\documentclass[a4paper,fleqn,usenatbib]{mnras}

\usepackage{newtxtext,newtxmath}

\usepackage[T1]{fontenc}
\usepackage{ae,aecompl}

\usepackage{graphicx} 
\usepackage{amsmath}  
\usepackage{ae,aecompl,graphicx,amsmath,bm,array,booktabs,mathtools,soul}
\usepackage{lipsum} 

\usepackage{hyperref}
\hypersetup{
  citecolor=cyan,      
}

\usepackage{graphicx}
\usepackage{color}
\definecolor{orange}{rgb}{1,0.5,0}
\definecolor{forestgreen}{rgb}{0.13, 0.55, 0.13}
\definecolor{bittersweet}{rgb}{1.0, 0.44, 0.37}
\definecolor{chartreuse}{rgb}{0.87, 1.0, 0.0}
\definecolor{darkorchid}{rgb}{0.6, 0.2, 0.8}

\newcommand{\cf}{cf.,~}
\newcommand{\ie}{i.e.,~}
\newcommand{\eg}{e.g.,~}
\newcommand*\diff{\mathop{}\!\mathrm{d}}

\title[Lattice-Boltzmann methods for radiative transport in computational
  astrophysics] {Beyond moments: relativistic Lattice-Boltzmann methods
  for radiative transport in computational astrophysics}

\author[Weih et al.]{L. R. Weih$^{1}$\thanks{weih@itp.uni-frankfurt.de}, 
A. Gabbana$^{2}$,
D. Simeoni$^{2,3,4}$,
L. Rezzolla$^{1,5,6}$,
S. Succi$^{7,8,9}$,
R. Tripiccione$^{2}$
\\
$^{1}$ Institut f\"ur Theoretische Physik, Max-von-Laue-Str. 1, 60438 Frankfurt,~Germany\\
$^{2}$ Universit\`a di Ferrara and INFN-Ferrara, I-44122 Ferrara,~Italy\\
$^{3}$ Bergische Universit\"at Wuppertal, D-42119 Wuppertal,~Germany\\
$^{4}$ University of Cyprus, CY-1678 Nicosia,~Cyprus\\
$^{5}$ School of Mathematics, Trinity College, Dublin 2, Ireland\\
$^{6}$ Helmholtz Research Academy Hesse for FAIR, Max-von-Laue-Str. 12, 60438
Frankfurt,~Germany\\
$^{7}$ Center for Life Nano Science @ La Sapienza, Italian Institute of Technology, Viale Regina Elena 295, I-00161 Roma,~Italy\\
$^{8}$ Istituto Applicazioni del Calcolo, National Research Council of Italy,
Via dei Taurini 19, I-00185 Roma,~Italy\\
$^{9}$ Harvard-SEAS, Oxford Street 29, 02130 Cambridge,~USA\\}

\date{Accepted XXX. Received YYY; in original form ZZZ}

\pubyear{2020}

\begin{document}
\label{firstpage}
\maketitle

\begin{abstract}
  We present a new method for the numerical solution of the
  radiative-transfer equation (RTE) in multidimensional scenarios
  commonly encountered in computational astrophysics. The method is
  based on the direct solution of the Boltzmann equation via an extension
  of the Lattice Boltzmann (LB) equation and allows to model the evolution
  of the radiation field as it interacts with a background fluid, via
  absorption, emission, and scattering. As a first application of this
  method, we restrict our attention to a frequency independent (``grey'')
  formulation within a special-relativistic framework, which can be
  employed also for classical computational astrophysics. For a number
  of standard tests that consider the performance of the method in
  optically thin, optically thick and intermediate regimes with a static
  fluid, we show the ability of the LB method to produce accurate and
  convergent results matching the analytic solutions. We also contrast
  the LB method with commonly employed moment-based schemes for the
  solution of the RTE, such as the M1 scheme. In this way, we are able to
  highlight that the LB method provides the correct solution for both
  non-trivial free-streaming scenarios and the intermediate optical-depth
  regime, for which the M1 method either fails or provides inaccurate
  solutions. When coupling to a dynamical fluid, on the other hand, we
  present the first self-consistent solution of the RTE with LB methods
  within a relativistic-hydrodynamic scenario. Finally, we show that
  besides providing more accurate results in all regimes, the LB method
  features smaller or comparable computational costs compared to the M1
  scheme. 

\end{abstract}

\begin{keywords}
  radiative transfer -- 
  radiation: dynamics -- 
  neutrinos -- 
  scattering -- 
  methods: numerical
\end{keywords}

\section{Introduction}
\label{sec:intro}

The proper treatment of the dynamics of radiation, as it interacts with a
matter fluid, is a fundamental problem in essentially all astrophysical
phenomena and requires the solution of the radiative-transport equation
(RTE). Given the complexity of the RTE and the nonlinear regimes normally
encountered in astrophysical and cosmological scenarios, the inclusion of
radiative effects inevitably commands the use of advanced numerical
methods to solve the RTE \citep[see, \eg][for the solution on an
  unstructured dynamic grid]{Paardekooper2010}.

An important and representative example is a binary system of merging
neutron stars (see \citet{Baiotti2016, Paschalidis2016} for an overview),
where the radiative transport of neutrinos can alter significantly the
chemical composition of the ejected matter, the efficiency of its
ejection, as well as the stability of the post-merger
object. Furthermore, radiative-transport effects are expected to play a
fundamental role in the kilonova signal that is produced
\citep{Rosswog2014a, Dietrich2016, Siegel2016a, Bovard2017, Perego2017,
  Fujibayashi2017b, Siegel2017, Fernandez2018} and might even be relevant
for producing a short gamma-ray burst associated with such a merger. An
equally important astrophysical scenario where radiation transport plays
a fundamental role is the one explored in simulations of neutrino-driven
core-collapse supernovae \citep{Mezzacappa01, Oconnor2015, Just2015b,
  Kuroda2016}, for which \citet{Colgate1966} first showed that the
radiation in form of neutrinos is essential for the explosion
mechanism. This is because the explosion mechanism may require a fine
balance, of the order of a few percent, between the energy deposited in
the stalled accretion shock and the release of potential gravitational
energy by the collapsing matter (see also \citet{Janka07} for an
overview). Finally, one more classical astrophysical scenario where
radiative-transfer effects cannot be ignored is the study of accretion
flows around black holes \citep{Zanotti2011, Fragile2012, Roedig2012,
  Sadowski2013, McKinney2014}, for which a broad array of techniques has
been developed over the years to compare with the observations
\citep{EHT_M87_PaperI}.

Unfortunately, the computational cost associated with the solution of the
RTE in numerical astrophysics also represents a significant obstacle to
the inclusion of radiative effects in numerical simulations. This is due
to the properties of the fundamental equation behind the RTE, \ie the
Boltzmann equation for massless particles, which lives in a
seven-dimensional space of time (one dimension), configuration space
(three dimensions) and momentum space (three more dimensions). As a
result, the solution of the RTE for typical astrophysical scenarios as
the ones mentioned above exceeds the current capacities of supercomputers
and thus approximate methods have to be used\footnote{In some cases the
  direct solution of the Boltzmann equation is indeed feasible by
  exploiting symmetries and reducing the spatial dimensionality of the
  problem.}.

A low-order approximation first used in the context of supernova explosions by
\citet{Livio1980} and commonly employed in binary neutron-star
simulations is the \textit{leakage scheme} \citep{Ruffert1996c,
  Rosswog:2003b, Galeazzi2013, Perego2014, Most2019b}, which only allows
for cooling via the emission of neutrinos and is therefore not useful for
core-collapse supernovae simulations, where heating is essential for
reviving the shock. These effects can be included by an approximation of
similar simplicity, the \textit{flux-limited diffusion} approximation first
implemented for astrophysical simulations of core-collapse supernovae by
\citet{Bruenn1975} \citep[see also][]{Pomraning1981, Levermore1981}; a recent implementation of this
scheme has been presented by \citet{Rahman2019}. In this method, the
zeroth moment of the radiation distribution function, \ie the radiation
energy-density, is evolved together with the fluid quantities. Since the
zeroth moment does not provide any information about the direction of the
radiation fluxes, its implementation is useful only for systems with
clear underlying symmetries. These symmetries are not present in the case
of binary neutron-star simulations, so the flux-limited diffusion does
not offer but a crude approximation of the radiative effects. This is
also true for the M0 scheme developed by \citet{Radice2016}, which also
evolves the lowest moment of the distribution function, but in the
free-streaming limit.

A considerably better approximation is the so-called truncated
moment-based scheme developed by \citet{Thorne1981} and first implemented
in general relativity by \citet{Rezzolla1994} in one dimension and by
\citet{Shibata2011, Cardall2013b} in three dimensions. Within these
schemes, the lowest moments up to order $N$ of the distribution function
are evolved, and the flux-limited diffusion method is then the limiting
case for $N=0$ of the general set of moment-based schemes. Increasing the
order of the hierarchy to the case with $N=1$ implies that the
momentum-density vector is evolved together with the radiation energy
density. Such a scheme is known in the literature as ``M1 scheme'' and is
indeed one of the most commonly used methods for radiative transport
throughout many different applications of computational relativistic
astrophysics \citep{Rezzolla1994, Roedig2012, Sadowski2013, Fragile2014,
  McKinney2014, Oconnor2015, Foucart2015a, Skinner2019, MelonFuksman2019,
  Weih2020b}. With this method, it is possible to track the
\textit{average} direction of the radiation momentum, providing a
significant improvement over the previously mentioned schemes, but can
still lead to rather unphysical results such as those that emerge when
radiation beams interact and cross \citep[see e.g.,][]{Fragile2014,
  Foucart2015a,Weih2020b}. Furthermore, as is typical in moment-based
schemes, in the M1 approach the set of evolution equations for the zeroth
and first moment depend on the second moment, which is not known within
this hierarchy scheme and has to be approximated in the form of a closure
relation. This results in the M1 scheme only being exact in the limits of
high and/or low optical depths, but not in the intermediate regime. The
situation does not improve when going to methods with $N>1$, which suffer
from increased computational costs and also need to specify a closure
relation that expresses the $N+1$ moment in terms of the lower-order
ones.

A more accurate solution of the RTE is offered by a completely different
class of methods employing Monte-Carlo techniques for the handling of the
radiation field \citep{Foucart2017,Miller2019a}, which, however, suffer
from low-statistics numerical noise and a comparatively high
computational cost.
To summarise, state-of-the-art numerical solutions of the RTE in
computational relativistic astrophysics revolve around two main classes
of methods: i) approximate methods based on the laws of hydrodynamics
(such as leakage, flux-limited diffusion, M1) or ii) direct solutions for
specific cases where symmetries can be exploited (such as Monte-Carlo
approaches to the solution of the Boltzmann equation).

The scope of this paper is to introduce a new method for the solution of
the RTE and hence for the treatment of radiative effects in computational
astrophysics that promises to be more precise than the M1 scheme,
providing a correct treatment of intersecting radiation beams and an
accurate treatment of regimes of high, low, and intermediate optical
depths. At the same time, it comes with an algorithmic complexity and an
associated computational cost comparable to that of M1 schemes, thus
making it well suited for multi-dimensional astrophysical simulations.

In essence, this new radiative-transport scheme stems from the
\textit{Lattice Boltzmann method} (LB method) \citep{krueger-book-2017,
  succi-book-2018}, which is commonly used in computational fluid
dynamics as an alternate scheme to direct hydrodynamic solvers.

The application of the LB method as a solver for radiation transfer
problems is relatively new \citep{asinari_2010_lbf}, and most models
proposed so far apply only to the analysis of steady-state
radiation-transport problems in one and two dimensions
\citep{bindra-pre-2012, mishra-hte-2014, mcculloch-cf-2016, yi-pre-2016},
with very few studies carried out in three dimensions \citep{McHardy2016,
  wang-pne-2019}.
More recent developments \citep{mink-jqsrt-2020} have shown in detail
that LB offers an accurate and efficient tool in the diffusive regime of
radiation transport, though struggling in the transition towards
ballistic conditions.
Finally, we should stress that all these previous studies deal with
radiative transfer in non-relativistic regimes.

We introduce a new LB solver for studying the time dependent evolution of
radiation that interacts via emission, absorption and scattering with a
(dynamic) background fluid. We make use of high-order spherical
quadrature rules, which, thanks to their high-order isotropy, allow to
significantly extend the applicability of the method to a wider range of
kinetic regimes. We work in a special-relativistic framework and present,
to the best of our knowledge, the first self-consistent coupled
simulation of an LB solver for radiative transport with a dynamically
evolving fluid background.

Our paper is structured as follows: in Sec. \ref{sec:LB} we present a
short summary and introduction of the classical LB method and discuss its
advantages, which make it ideal for radiative-transport problems in
computational astrophysics. In Sec. \ref{sec:methods} we illustrate the
details of the new LB method for radiative transport in special
relativity and within a ``grey'' (\ie energy averaged) approximation and
show how to implement such a scheme suitably for simulations. We verify
this implementation by a number of standard tests in Sec.
\ref{sec:tests}. In Sec. \ref{sec:coupled} we couple our new LB-code to
a hydrodynamics code, which is representative of the many
relativistic-hydrodynamics codes used in the field of numerical
astrophysics, and present a simulation of a relativistic jet; we show
that its dynamics changes qualitatively due to the back-reaction of the
produced radiation. Finally, we compare its accuracy and computational
cost to the commonly used M1 scheme in Sec. \ref{sec:benchmarks}. We
conclude and discuss future prospects of our new method in
Sec. \ref{sec:outlook}.

Throughout this paper, we use units with $c=1$ and only write the speed
of light explicitly in equations where it is necessary for clarity. We
write three-vectors in boldface while unit-vectors carry a hat.

\section{A short introduction to Lattice Boltzmann}
\label{sec:LB}

In this section we provide a brief overview of the LB method. The reader
already familiar with the topic may safely jump directly to
Sec.~\ref{sec:methods}.
We should stress that, for ease of presentation, in this section we will
summarise the conceptual steps of the derivation and algorithmic
structure of LB in a non-relativistic framework. For details on the
derivation of the method in special-relativity, the interested reader is
referred to a recent review \citep{gabbana-pr-2020}.

The LB method has emerged in the past decades as a computationally
efficient numerical tool for the simulation of the dynamic of fluids in
classical hydrodynamics. Its origin can be traced back to the pioneering
work on discrete velocity models \citep{broadwell-pf-1964} in the 1960s
and later on to the work done on Lattice Gas Cellular Automata
\citep{hardy-prl-1973, frisch-prl-1986} in the late 1980s. Since then,
the method has evolved as an independent and efficient alternative to
direct Navier-Stokes solvers in the field of classical computational
fluid dynamics \citep{mcnamara-prl-1988, higuera-epl-1989, Benzi1992} and allied
disciplines, primarily soft matter
\citep{succi-epl-2015,duenweg-book-2009}.

Contrary to direct hydrodynamic solvers, the LB method relies on the
underlying microscopic dynamics of the fluid constituents -- be them
molecules or photons -- and therefore the natural theoretical framework
to start approaching the method is kinetic theory and its mathematical
cornerstone, the Boltzmann equation (here taken without source terms):
\begin{align}\label{eq:boltzeq}
  \left( \frac{\partial }{\partial t} + \bm{\varv} \cdot \nabla \right)
  f(\bm{r},\bm{\varv},t) = \mathcal{C}(\bm{r},\bm{\varv},t)   \, .
\end{align}
The distribution function $f(\bm{r},\bm{\varv},t)$ refers to the number of
particles with velocity $\bm{\varv}$ at position $\bm{r}$ at time $t$,
while the collision operator $\mathcal{C}(\bm{r},\bm{\varv},t)$ accounts
for collisions between point-particles in the fluid and in Boltzmann's
theory takes the form of a non local integral in momentum space.
It is customary to replace the full collisional operator with a
simplified model, such as the well-known Bhatnagar-Gross-Krook (BGK) relaxation
time approximation \citep{bhatnagar-pr-1954}, encompassing the natural
tendency of the system to relax towards an equilibrium, \ie the tendency
of $f$ to reach a distribution function $f^{\rm{eq}}$ describing a local
equilibrium state
\begin{align}\label{eq:bgk}
  \mathcal{C}(\bm{r},\bm{\varv},t) = - \frac{1}{\tau}(f(\bm{r},\bm{\varv},t) -
	f^{\rm{eq}}(\bm{r},\bm{\varv},t))  \, .
\end{align}
In the above, $\tau$ represents the typical timescale needed to reach the
equilibrium, a parameter which controls the hydrodynamic transport
coefficients, hence dissipative phenomena within the fluid.

For a classical fluid with massive constituents, the equilibrium function
is represented by the Maxwell-Boltzmann distribution \citep[see,
  \eg][]{Rezzolla_book:2013}
\begin{align}\label{eq:feq}
  f^{\rm{eq}}(\bm{r},\bm{\varv},t) = \rho (\bm{r}, t) \left( \frac{m}{2 \pi k_{\rm B} T}\right)^\frac{d}{2}
	\mathrm{exp}\left[-\frac{m}{k_{\rm B} T}(\bm{\varv}-\bm{u}(\bm{r},t))^2\right] \, , 
\end{align}
where $m$ is the mass of constituent particles and $k_{\rm B}$ is the
Boltzmann constant. The rest-mass density $\rho(\bm{r},t)$ and the
velocity field $\bm{u}(\bm{r},t)$ can be computed as the zeroth and first
moment of the distribution function, respectively,
\begin{align}\label{eq:moments}
  \rho (\bm{r}, t) := m \int f(\bm{r}, \bm{\varv}, t) d \bm{\varv}    \, ,
  \quad\quad 
  \bm{u} (\bm{r}, t) := \frac{m}{\rho (\bm{r}, t)} \int \bm{\varv} f(\bm{r}, \bm{\varv},
  t) d \bm{\varv}     \, .
\end{align}

Historically, the LB method was devised as a noise-free (pre-averaged)
version of its lattice-gas cellular automaton (LGCA) ancestor
\citep{mcnamara-prl-1988}. This represented a major conceptual leap, but
left all other LGCA shortcomings untouched, primarily the exponential
complexity barrier associated with Boolean collision operators and the
ensuing low collisional rates which prevented LGCA from accessing
high-Reynolds regimes (turbulent flows). All of the above barriers were
lifted just months later in a short sequence of papers
\citep{higuera-epl-1989, higuera-epl-1989b,higuera-epl-1989c, Benzi1992},
which placed LB on the map of computational fluid dynamics.

For the sake of simplicity and continuity with continuum kinetic theory,
it proves expedient to derive LB from the expansion of the equilibrium
distribution in a series of orthogonal Hermite polynomials
$\bm{H}^{(k)}(\bm{\varv})$; the advantage of using Hermite as the
expansion basis is that the expansion coefficients $\bm{a}^{(k)}$
coincide with the moments of the distribution function
\citep{grad-cpam-1949b, grad-cpam-1949}. The expansion is then truncated
to the desired order $S$, high enough to recover the macroscopic
observables of interest \citep{shan-prl-1998,shan-jofm-2006}.

In $d$ dimensions, this results in the following local equilibrium
distribution function:
\begin{align}\label{eq:trunc_feq}
	f^{\rm eq} = \left(\frac{1}{2\pi} \right)^\frac{d}{2}
	\mathrm{exp}\left(-\frac{\bm{\varv}^2}{2}\right) \sum_{k=0}^{S} \bm{a}^{(k)}(\rho, \bm{u}) \bm{H}^{(k)}(\bm{\varv}) \,, 
\end{align}
where all physical quantities have been made dimensionless by appropriate
scaling with respect to a characteristic velocity $\tilde{c}:=
\sqrt{k_{\rm B} T_0 / m_0 }$, a characteristic temperature $T_0$, a mass
unit $m_0$, and a length scale $L_0$.
Furthermore, the (microscopic) velocity vector of  phase-space is
discretized using a set of $N_{\mathrm{pop}}$ distinct populations
$\bm{\varv}_i$ with $i\in [1;N_{\mathrm{pop}}]$. As a consequence, the
distribution $f$ itself becomes a set of $N_{\mathrm{pop}}$ functions
$f_i(\bm{r},t)=f(\bm{r},\bm{\varv}_i,t)$, each accounting for the
particles moving along the discrete direction $\bm{\varv}_i$.
  
The choice of the discrete velocities lies at the heart of the LB method, the
informing criterion being of reproducing {\it exactly} the set of kinetic moments
which describe the low-Knudsen hydrodynamic regime, namely the mass density 
(scalar), the flow current (vector) and the momentum flux tensor (second-order tensor).
 In the above, "exactly" means that no error altogether is incurred by replacing the
 integrals in continuum velocity space with the corresponding summations over the
 discrete velocities which characterise the LB representation. 
Formally, this can be linked to a Gauss-Hermite quadrature rule, where one defines
$\bm{\varv}_i$, and the corresponding weights $w_i$, and requires
\textit{exact} preservation of the relevant hydrodynamic fields.
In equations:
\begin{align}\label{eq:disc_moments}
  \rho (\bm{r}, t) &= \sum_{i=1}^{N_{\mathrm{pop}}} f_i(\bm{r}, t)    \, , 
  \quad\quad 
  \bm{u} (\bm{r}, t) = \frac{1}{\rho (\bm{r}, t)} \sum_{i=1}^{N_{\mathrm{pop}}} \bm{\varv}_i f_i(\bm{r}, t)      \, ,
\end{align}
with the truncated equilibrium distribution Eq. (\ref{eq:trunc_feq}) given by
\begin{align}\label{eq:disc_feq}
  f_i^{\rm eq} = w_i \sum_{k=0}^{S} \bm{a}^{(k)}(\rho, \bm{u}) \bm{H}^{(k)}(\bm{\varv}_i)  \, .
\end{align}
The combination of the velocity discretization with explicit time-marching
finally delivers the lattice Boltzmann equation:
\begin{equation}\label{eq:LB_classic}
  f_i(\bm{r} + \bm{\varv}_i \Delta t, t+\Delta t) = f_i(\bm{r}, t) +  \Delta
  t \mathcal{C}_i(\bm{r}, t)\, ,
\end{equation}
with $\Delta t$ the time-step, $\Delta x = \bm{\varv}_i \Delta t$ the
characteristic mesh spacing, and 
\begin{equation}\label{eq:bgk_discrete}
  \mathcal{C}_i = \frac{1}{\tau} (f_i - f_i^{\rm eq}) .
\end{equation}
The evolution of Eq. (\ref{eq:LB_classic}) follows the so-called
``stream-and-collide'' paradigm, where in the ``collide step'' each
population $f_i(\bm{r}, t)$ is updated by receiving a local collisional
contribution:
\begin{equation}\label{eq:collide}
  f_i^*(\bm{r}, t) = f_i(\bm{r}, t) + \Delta t \mathcal{C}_i(\bm{r}, t)      \, .
\end{equation}
In the streaming step, instead, the post collision populations $f_i^*(\bm{r},
t)$ stream along their associated direction $\bm{\varv}_i$, landing
on the corresponding  neighbouring lattice site (no particle can fly off-grid):
\begin{equation}\label{eq:streaming}
  f_i(\bm{r} + \bm{\varv}_i \Delta t, t+\Delta t) = f_i^*(\bm{r}, t) \, .
\end{equation}

Two major assets associated to the stream-collide paradigm are worth
highlighting.  First, the non-local operator (streaming) is linear and the
nonlinear one (collision) is local, meaning that, at variance with the
hydrodynamic representation, non-linearity and non-locality are
disentangled. This is because information always travels along constant
characteristics, the discrete velocities, regardless of the spacetime
complexity of the emergent hydrodynamics.  By contrast, in the fluid
representation, information travels along space-time dependent material
lines, defined by the local flow speed.  This is a major advantage also
for the handling of complex boundary conditions and parallel computing.

Second, since dissipation emerges from enslaving of the Boltzmann
distribution to local equilibrium, there is no need for second order
spatial derivatives.  This is a significant advantage for the calculation
of the stress tensor, especially near solid boundaries.  In addition,
since the collision operator is conservative to machine-accuracy, the LB
method usually offers better accuracy than most grid-based
discretisations of the Laplace operator.

The standard LB method described so far is suitable for the description
of hydrodynamic systems, where the molecular mean free path is much
shorter than the shortest hydrodynamic length scale, the ratio of the two
being the Knudsen number of the fluid.
On the other hand, for a fluid of radiation particles, \eg
photons or neutrinos, one should bear in mind that the radiation
constituents do not interact among themselves, but only with the
background fluid, which effectively produces, destroys, and scatters the
radiation particles. Hence, when applying the LB method to radiation
local conservation laws must be revisited and complemented with suitable 
source (sink) terms, accounting for the above processes.
In particular, energy and momentum are conserved only over the
combined system of radiation and fluid.

This still fits within the LB framework, which has been used for decades to
study transport phenomena, such as advection-diffusion-reaction equations,
\citep{massaioli-epl-1993,he-jocp-1998, peng-pre-2003, karlin-pre-2013}.
Based on this idea, several LB models have been proposed to study
radiative transport. Initially most of these efforts have focused on
studying steady-state problems in one and two spatial dimensions
\citep{asinari_2010_lbf, bindra-pre-2012, mishra-hte-2014,
  mcculloch-cf-2016, yi-pre-2016}, considering isotropic as well as
anisotropic scattering \citep{vernekar-ijhmt-2014}. So far, however, only
very few authors have considered the three-dimensional case
\citep{McHardy2016, mink-jqsrt-2020}.

All the models mentioned above discretize the velocity space by means of
standard space-filling lattices, typically used in LB methods. As a
consequence, information travels in a single timestep to nodes located at
different distances, corresponding to a different magnitude of the
discrete velocities, a typical example being the two-dimensional
nine-velocity lattice comprising a rest particle with zero speed, four
particles connecting to the nearest neighbours (speed $1$) and four
connecting the diagonals (speed $\sqrt 2$). The latter are mandatory to
the correct recovery of the two-dimensional Navier-Stokes equations, but
not necessary (albeit recommended for matters of accuracy) for the case
of advection-diffusion equations, a property which can be transferred to
radiative LBEs \citep{gairola-ane-2017, wang-pne-2019}.
The obvious drawback is that such models are appropriate only for
collision-dominated, low-diffusive, regimes, while higher-order models
based on extended velocity sets are required to describe the transition
from low to high diffusivity and finally towards the free-streaming
(ballistic) regimes.

In the following, we introduce a new LB method which is precisely meant
to address this important issue. In particular, we employ high-order
spherical quadratures to provide a unified numerical RTE solver, capable
of handling both low and high diffusive regimes, up to the free-streaming
scenario.

This major extension comes at a price: since the discrete velocities lie
on a sphere, they no longer end on the nodes of a space-filling Cartesian
grid. Hence interpolation is required to supplement the standard
stream-and-collide algorithm, leading to the loss of exact streaming.
Actual simulations show that in all cases inspected in this paper, the
lack of exact streaming does not lead to any appreciable loss of accuracy
of the method, thus identifying the radiative LB developed here as a viable
and competitive numerical RTE solver. 

\section{The Lattice Boltzmann scheme for radiative transport}
\label{sec:methods}

\subsection{Mathematical setup}
\label{sec:methods1}

The RTE is the master equation describing how radiation propagates
through a medium that scatters, absorbs and emits radiation particles. As
such, the RTE follows from the Boltzmann equation assuming massless
particles (\ie photons or neutrinos).
All radiative fields are expressed in terms of the distribution function
$f_{\nu}(\bm{x},\bm{\hat{n}},\nu)$ of neutrinos or photons within a given
frequency band $\diff \nu$ at position $\bm{x}$ and velocity within a
solid angle $\diff \Omega$ in the direction $\bm{\hat{n}}$.

The subscript $\nu$ highlights the dependency on frequency, to
distinguish from the corresponding frequency-independent quantities to be
introduced later on.
The RTE describes the evolution of the radiation distribution
function in the direction $\bm{\hat{n}}$ 
\begin{equation}\label{eq:rte_nu}
  \frac{1}{c}\frac{\partial f_{\nu}}{\partial t} + \bm{\hat{n}} \cdot \nabla f_{\nu}
  =
  - \kappa_{a,\nu} f_{\nu} + \eta_{\nu} 
  + \mathcal{C}_\mathrm{scat} =: \mathcal{C}_\mathrm{rad}\,.
\end{equation}
This expression is equivalent to Eq. (\ref{eq:boltzeq}), but for massless
particles and with an explicit expression for the collisional operator
$\mathcal{C}_\mathrm{rad}$ on the right-hand side. This operator
splits into three terms: the absorption, the emission, and the scattering
term, respectively. The absorption term is proportional to the absorption
coefficient $\kappa_{a,\nu}$, the emission term to the emissivity
$\eta_\nu$, and the scattering term $\mathcal{C}_\mathrm{scat}$ is
written in its most general form as \citep{Bruenn85,Rampp97}
\begin{align}
\label{eq:scattering}
  \mathcal{C}_\mathrm{scat} =  \int_0^\infty \nu^{\prime 2} \diff \nu^\prime
  \int_{4\pi} f^\prime_{\nu^\prime} (1-f_\nu) R^{\mathrm{in}}
    -f_\nu (1-f^\prime_{\nu^\prime})R^{\mathrm{out}} \diff \Omega^\prime
   \, , \nonumber \\
\end{align}
where $R^{\mathrm{in}}(\nu, \nu^\prime)$ and $R^{\mathrm{out}}(\nu,
\nu^\prime)$ are the incoming and outgoing scattering kernels,
respectively. 

These kernels depend on the underlying physical process and generally do
not allow for an analytic solution of the scattering integral. In order
to simplify the above integral, the incoming and outgoing scattering
kernels are typically expanded as a Legendre series and truncated to the
first two terms \citep{Bruenn85, Rampp97,Shibata2011}. This leads to
\begin{equation}
\label{eq:kernel}
  R^{\mathrm{in/out}}(\nu,\nu^\prime) \approx \frac{1}{2}
  \Phi_0^{\mathrm{in/out}}(\nu,\nu^\prime) + \frac{3}{2}
  \Phi_1^{\mathrm{in/out}}(\nu,\nu^\prime)\, \mathrm{cos}\,\theta \, ,
\end{equation}
where $\Phi_{\ell=0,1}^{\mathrm{in/out}}$ are the $\ell$-th coefficients
of the Legendre expansion and $\theta$ is the angle between the incoming
and outgoing particle, $\mathrm{cos}\, \theta = \bm{\hat{n}} \cdot
\bm{\hat{n}^\prime}$.

Hereafter, we will consider only iso-energetic scatterings, \ie we assume
that the energy of the radiation particles is left unchanged by the
scattering with the constituents of the underlying fluid.

It follows that:
$\Phi_{\ell}^{\mathrm{in}}(\nu, \nu^\prime) \equiv
\Phi_{\ell}^{\mathrm{out}}(\nu,\nu^\prime) \eqqcolon
\Phi_{\ell}(\nu)\delta(\nu-\nu^\prime)$.

In this way, inserting Eq. (\ref{eq:kernel}) in
Eq. (\ref{eq:scattering}), the scattering term in Eq. (\ref{eq:rte_nu})
reads as:
\begin{align}\label{eq:expanded-scattering-term}
  \mathcal{C}_{\mathrm{scat}} &\approx - \kappa_{0,\nu} f_\nu 
     + \kappa_{0,\nu} E_\nu + 3\kappa_{1,\nu}\bm{\hat{n}}\cdot\bm{F}_\nu
  \, ,
\end{align}
where we have used the definition of the zeroth and first moment of the
distribution function
\begin{align}
  E_\nu &\coloneqq \frac{1}{4\pi} \int_{4\pi} f_\nu \diff \Omega \, ,
  \quad\quad \bm{F}_\nu \coloneqq \frac{1}{4\pi} \int_{4\pi} \bm{\hat{n}}f_\nu \diff \Omega \, ,
\end{align}
and defined the energy-dependent opacities $\kappa_{\ell,\nu} =
2\pi\nu^2\Phi_{\ell}$. Note that the explicit form of these opacities
depends on the type of radiation, namely, whether one is considering
photons or neutrinos (see \eg Eqs. (A.47) and (A.48) in \citet{Rampp97}
for the case of neutrinos or \citet{Rybicki_Lightman1986} for photons).

The frequency-integrated version of the RTE -- often referred to as
``grey'' approximation -- is obtained via multiplication by $\nu^3$ and
integration over $\nu$, \ie
\begin{equation}\label{eq:rte_gray}
  \frac{1}{c}\frac{\partial I}{\partial t} + \bm{\hat{n}} \cdot \nabla I
  =
  - \kappa_{a} I + \eta + \kappa_{0}(E-I) + 3\kappa_{1}\bm{\hat{n}} \cdot \bm{F}
  \,,
\end{equation}
where we used the definition of the frequency-integrated specific
intensity
\begin{equation}\label{eq:I}
  I \coloneqq \int_0^\infty I_\nu \diff \nu \coloneqq \int_0^\infty \nu^3 f_{\nu} \diff \nu\,,
\end{equation}
and the frequency-integrated moments
\begin{align}
  E &\coloneqq \int_0^\infty \nu^3 E_\nu \diff \nu = \frac{1}{4\pi}
  \int_{4\pi} I \diff \Omega \,, \label{eq:rad_moments_E} \\
  \bm{F}
  &\coloneqq \int_0^\infty \nu^3 \bm{F}_\nu \diff \nu = \frac{1}{4\pi}
  \int_{4\pi} \bm{\hat{n}} I \diff \Omega \label{eq:rad_moments_F} \, .
\end{align}
These moments can be interpreted as the radiation energy density and
momentum density, respectively, and arguably represent the most important
properties of the radiation field.

Finally, the energy-averaged opacities $\kappa_a$, $\kappa_0$, $\kappa_1$
are given by
\begin{equation} \label{eq:kappa}
  \kappa_{\ast} \coloneqq \frac{\int_0^\infty \kappa_{\ast, \nu} I_\nu \diff
    \nu} {\int_0^\infty I_\nu \diff \nu} \,,
\end{equation}
where $\ast=a, 0, 1$ and the
frequency-integrated emissivity $\eta$ is given by
\begin{equation}
  \label{eq:eta}
  \eta \coloneqq \int_0^\infty \nu^3 \eta_\nu \diff \nu \,.
\end{equation}
%

\subsection{LB discretization of the RTE}
\label{sec:methods2}

We next describe the steps needed to derive an LB-inspired discretization
of the RTE within the grey approximation. We start by recasting
Eq. (\ref{eq:rte_gray}) in a BGK-like form:
\begin{align}\label{eq:rte_gray_bgkform}
  \frac{1}{c}\frac{\partial I}{\partial t} + \bm{\hat{n}} \cdot \nabla I
  =
  - \kappa_{0} \left( I - I^{\rm eq} \right) + S \,,
\end{align}
where the source term $S$, accounting for emission and absorption, 
is given by
\begin{align}
 \label{eq:rte_gray_bgkform_2}
S = - \kappa_{a} I + \eta\,,
\end{align}
while the scattering term has been rearranged by introducing the
equilibrium radiation intensity
\begin{align}
  I^{\rm eq} := E + \lambda \bm{\hat{n}} \cdot \bm{F} , \quad  \lambda = 3 \frac{\kappa_{1}}{\kappa_{0}} \,.
\end{align}

We then perform a discretization of the velocity space, namely, the
directions $\hat{\bm{n}}$ along which radiation propagates at the speed
of light. In essence, we replace $\hat{\bm{n}}$ with a discrete set of
$N_{\mathrm{pop}}$ directions $\hat{\bm{n}}_i$, which define the
corresponding discrete intensities $I_i(\bm{x},t) =
I(\bm{x},\hat{\bm{n}}_i,t)$ and their associated weights $w_i$ (a more
detailed description on how to choose the $N_{\mathrm{pop}}$ directions
is given in Sec \ref{sec:methods2.2}). Consequently,
Eq. (\ref{eq:rte_gray_bgkform}) splits into a set of $N_{\mathrm{pop}}$
equations, each describing the evolution of a specific intensity $I_i$
via the $i$-th expression of the RTE
\begin{align}\label{eq:rte_gray_vel_discrete}
  \frac{1}{c}\frac{\partial I_i}{\partial t} + \bm{\hat{n_i}} \cdot \nabla I_i
  =
  - \kappa_{0} \left( I_i - I_i^{\rm eq} \right) + S_i
  \,.
\end{align}
The discrete counterparts of the integrals in
Eq. (\ref{eq:rad_moments_E}) and Eq. (\ref{eq:rad_moments_F}) then take
the following form:
\begin{align}\label{eq:moments_discrete}
  E \approx \sum_{i=1}^{N_{\mathrm{pop}}} I_i \,, \quad\quad 
  \bm{F} \approx \sum_{i=1}^{N_{\mathrm{pop}}} \bm{\hat{n}}_i I_i  \,,
\end{align}
while higher moments can be computed accordingly as
\begin{equation} \label{eq:higher_moments}
	M^{j_1 \dots j_m} \approx \sum_{i=1}^{N_\mathrm{pop}} n_i^{j_1}\dots
	n_i^{j_m} I_i \, .
\end{equation}

Finally, a first-order discretization in time with a time-step $\Delta t$
leads to the radiative Lattice Boltzmann equation:
\begin{align}\label{eq:rte_discrete}
  I_i(\bm{r} + c\bm{\hat{n}}_i\Delta t, t + \Delta t) &= I_i(\bm{r},t)- c
	\kappa_0 \Delta t \left( I_i(\bm{r},t) - I_i^{\rm eq}(\bm{r},t) \right)                              \notag \\
    & + c \Delta t S_i(\bm{r},t) \, , 
\end{align}
where
\begin{align}\label{eq:source-and-equil}
  I^{\rm eq}_i := w_i \left( E(\bm{r},t) + \lambda \bm{\hat{n}_i} \cdot
  \bm{F}(\bm{r},t) \right)\,,
  \quad  
  S_i := w_i \eta - \kappa_a  I_i(\bm{r},t)\,.
\end{align}

\subsubsection{Velocity Discretization}
\label{sec:methods2.2}

The key of the LB method lies within the velocity discretization, that
is, the mapping of the continuum velocity space in terms of a finite set
of discrete velocities $\{ {\bm{v}}_i \}$, the so-called discrete
velocity stencil.

Indeed, the definition of the discrete velocity set, and its associated
weights, plays a crucial role in the LB method, and several approaches
have been developed over the years.
The one based on the Gauss-Hermite quadrature is arguably the most
systematic \citep{shan-jofm-2006,philippi-pre-2006, shan-jocs-2016}.
Another way to go, which has been used in the early days of LB,
is to construct velocity sets as $d$-dimensional projections
from known $(d+1)$ velocity sets \cite{humieres-epl-1986}.

Yet, another possible approach consists in defining general conditions
that should be satisfied by the velocity set, in terms of symmetry and
conservation, typically mass, momentum and momentum flux (isotropy).

As mentioned above, since we cannot rely on conservation laws that would
allow us to derive quadrature rules in a systematic way, we need to
define the velocity stencil on the basis of symmetry considerations and
isotropy conditions, privileging those stencils that exhibit a
sufficiently high order of isotropy, so as to handle the different
kinematic regimes that are typically encountered in astrophysical
scenarios.

Formally, we define $k$-rank tensors $T^{\alpha_1 \dots \alpha_k}$, that
are constructed by combining all the products between the different
directions $n_i$ forming the velocity stencil (appropriately weighted
with the weights $w_i$)
\begin{align}\label{eq:iso-tensors}
  T^{\alpha_1 \dots \alpha_k} \coloneqq \sum_i w_i n_i^{\alpha_1} \dots n_i^{\alpha_k}    \, .
\end{align}
The microscopic relations that have to be satisfied by the lattice in order to
ensure n-th order isotropy are given by \citep{rivet-book-2001}
\begin{align}\label{eq:isotropy-cond}
  T^{\alpha_1 \dots \alpha_k} 
  \begin{cases}
    =0                                                                                                             &\text{k odd} \hfilneg    \, ,\\
    \propto \sum\limits_{\text{perm}} \left(\delta_{\alpha_1\alpha_2}\dots\delta_{\alpha_{k-1}\alpha_{k}} \right) &\text{k even}            \, ,
  \end{cases}
\end{align}
which need to be satisfied for all $k \leq n$. 

One additional condition on the definition of the velocity stencil is
that all the (pseudo)-particles travel at the same speed, \ie the speed
of light, so $\bm{v}_i = c~\hat{\bm{n}}_i$. It follows that the discrete
directions $\hat{\bm{n}}_i$ must have the same magnitude, hence span the
surface of a sphere in three dimensions (3D), or a circle in two
dimensions (2D).
  
The adoption of spherical velocity stencils is not new to LB: it has been
used, for example, in LB models for the simulation of ultra-relativistic
hydrodynamics. In such frameworks, since the interest is restricted to
the hydrodynamic picture, it is still possible to define stencils which
live on the intersection between a Cartesian grid and a sphere of fixed
radius \citep{mendoza-prd-2013,gabbana-cf-2018}, thus preserving a very
desirable property of LB: exact-streaming.
On the other hand, going beyond hydrodynamics generally requires a much
larger number of discrete directions, making the definition of on-lattice
quadratures impractical. In these cases, it is therefore necessary to
take into consideration off-lattice schemes
\citep{coelho-cf-2018,ambrus-prc-2018}.
Since we need to properly model free-streaming regimes, we adopt this
latter approach and work with off-lattice stencils spanning a unit
sphere.

Moreover, the choice of the velocity set should be such to maximise the
accuracy in the calculation of the moments of the specific intensity 
$I(\bm{x},\bm{\hat{n}},t)$, such as the energy density $E(\bm{\hat{x}},t)$ 
and the momentum density $\bm{F}(\bm{\hat{x}},t)$.

In practice, we request the discrete sums in
Eq. (\ref{eq:moments_discrete}) to correctly reproduce their continuous
counterparts, \ie Eq. (\ref{eq:rad_moments_E}) and
Eq. (\ref{eq:rad_moments_F}). These are spherical integrals of the form
\begin{equation}\label{eq:sph-integral}
  Q(h) = \frac{1}{4\pi} \int_{4\pi} h(\bm{\hat{n}}) d \Omega \approx \sum_{i =
	1}^{N_{\mathrm{pop}}} w_i h(\bm{\hat{n}_i})  \, ,
\end{equation}
where $h(\bm{\hat{n}})$ represents a generic function of the direction
$\bm{\hat{n}}$. To begin with, we recall that any square integrable function
can be expanded on the unit sphere as a series of orthogonal
spherical harmonics \citep{atkinson-book-2012}
\begin{equation}\label{eq:harmonic-exp}
  h(\bm{\hat{n}}) = h(\theta, \phi) = \sum_{\ell = 0}^{+\infty} \sum_{m =
	-\ell}^{\ell} c_{\ell m} Y_\ell^m (\theta, \phi)  \, ,
\end{equation}
where the convergence rate depends on the coefficients $c_{\ell m}$. The
spherical quadrature rule determining the weights $w_i$ and discrete
directions $\bm{\hat{n}_i} = (\theta_i, \phi_i)$ is then said to be of
order $p$ if it integrates exactly all the spherical harmonics $Y_\ell^m
(\theta, \phi)$ up to the degree $\ell = p$
\begin{align}\label{eq:sph-quad-cond}
  \frac{1}{4\pi}\int_{4\pi} Y_\ell^m(\theta, \phi) d \Omega = \sum_{k =
	0}^{N_{\rm pop}-1} w_k Y_\ell^m(\theta_k, \phi_k) \quad\quad \forall\ \ell \leq p \,.
\end{align}
It follows that a spherical quadrature of order $p$ integrates exactly
all integrals $Q(h)$ of functions $h(\theta, \phi)$ described by linear
combinations of the first $p$ harmonics. On the other hand, functions
$h(\theta, \phi)$ that contain in their series harmonics of higher order
are only approximated by the quadrature, with errors that depend on the
smoothness of the function itself. Note that it is possible to prove that
quadratures of order $p$ satisfy Eq. (\ref{eq:isotropy-cond}) up to the
level $p$, enabling in this way to evaluate the quality of the stencils
with the use of a single parameter, \ie $p$ that we call the quadrature
order.

\begin{figure}
  \includegraphics[width=1.0\columnwidth]{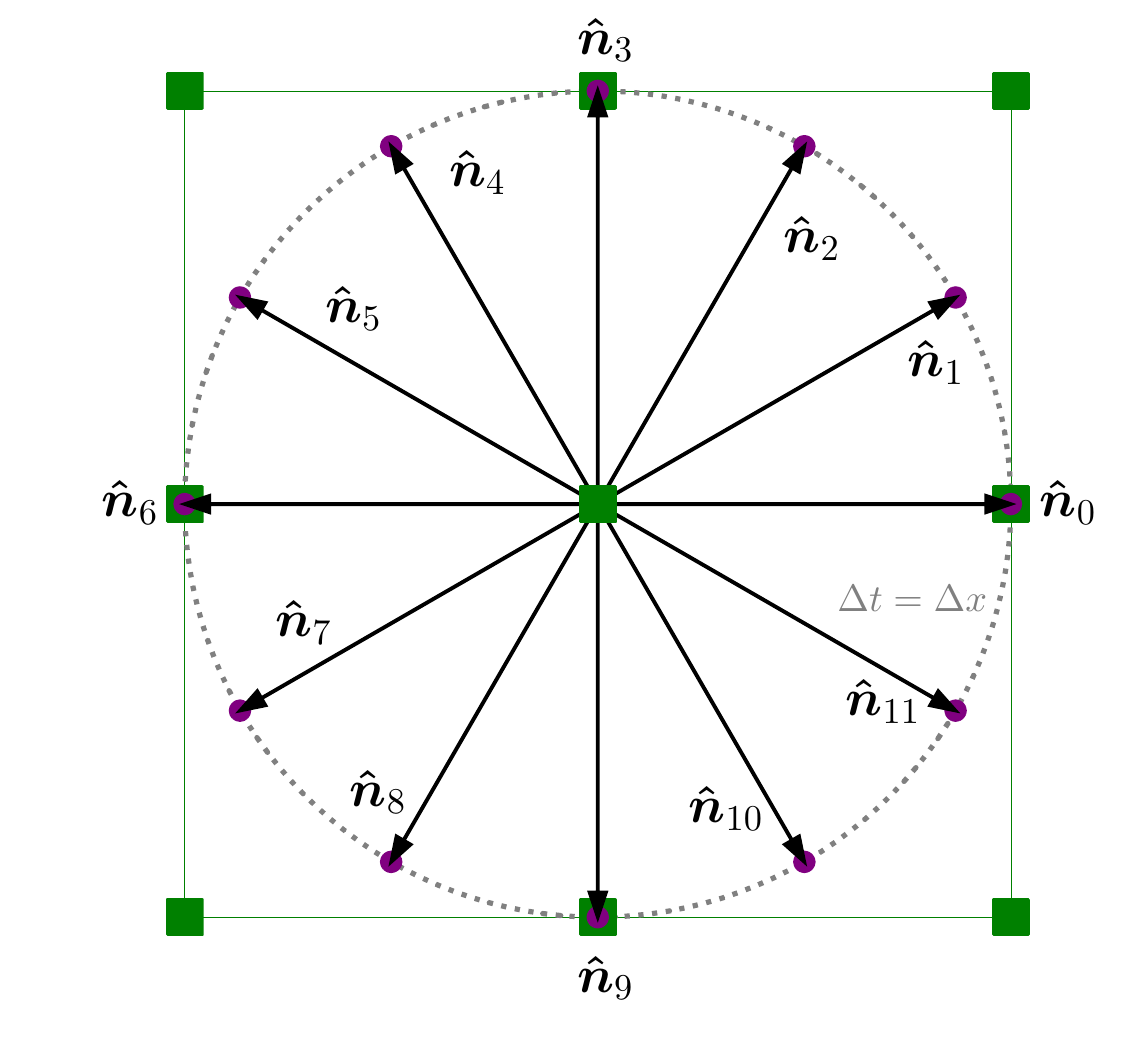}
  \caption{Example of two-dimensional velocity stencils showing
    $N_{\mathrm{pop}}=12$ discrete velocity directions (purple dots), in
    which the radiation is allowed to propagate. Green squares indicate
    the grid's cell-centres. The arrows end on a circle with radius
    $c\Delta t$, where $\Delta t$ is assumed to be in units of the
    grid-spacing.}
  \label{fig:stencil2D}
\end{figure}

Determining the quadrature satisfying the conditions discussed above is
trivial in the case of two dimensions: the integral in
Eq. (\ref{eq:sph-integral}) simply has to be computed on the unit
circle. Furthermore, the expansion in Eq. (\ref{eq:harmonic-exp}) reduces
to a Fourier series since the three-dimensional spherical harmonics
reduces to circular functions (sine and cosines of the only angular
coordinate $\phi$)
\begin{equation}\label{eq:harmonic-exp-2d}
  h(\bm{\hat{n}}) = h(\phi) = \sum_{m = -\infty}^{\infty} c_{m} e^{im\phi}  \, ,
\end{equation}
and one has to satisfy exactly the relations
\begin{align}\label{eq:sph-quad-cond-2d}
  \frac{1}{2\pi}\int_0^{2\pi} e^{im\phi} d \phi = \sum_{k =
    0}^{N_{\mathrm{pop}}-1} w_k e^{im\phi_k} \quad\quad \forall\ |m| \leq
  p \,.
\end{align}
As a result, to obtain a 2D quadrature of order $p$, \ie in order to
satisfy Eq. (\ref{eq:sph-quad-cond-2d}), it is sufficient to consider
$N_{\mathrm{pop}} = p + 1$ uniformly spaced points on the unit circle,
displaced by the angles $\phi_i = 2\pi\,k/N_{\mathrm{pop}}$, thus
$\bm{\hat{n}}_i =
\left[\mathrm{cos}(\phi_i),\,\,\mathrm{sin}(\phi_i)\right]^T$, all having
equal weights $w_k = 1 / N_{\mathrm{pop}}$. A representation of such a
velocity stencil with its discrete directions $\bm{\hat{n}}_i$ is
presented in Fig. \ref{fig:stencil2D}.

The problem becomes considerably more involved for the three-dimensional
case, which is clearly the most relevant one in terms of astrophysical
applications. Indeed, the definition of quadrature rules on the surface
of a sphere (\ie a 2-sphere) is still an active area of research, with
several different approaches coming with corresponding advantages and
drawbacks, depending on the target application
\citep{beentjes-arxiv-2015,gamba-josc-2017, gross-jocp-2018,
  lutsko-pre-2018, jiri-aa-2020}.

Here, we consider three different types of spherical quadrature schemes:
\begin{enumerate}

\item ~~Gauss-Legendre quadrature

\item ~Lebedev quadrature

\item Spherical design

\end{enumerate} 

\begin{figure*}
  \includegraphics[width=2.0\columnwidth]{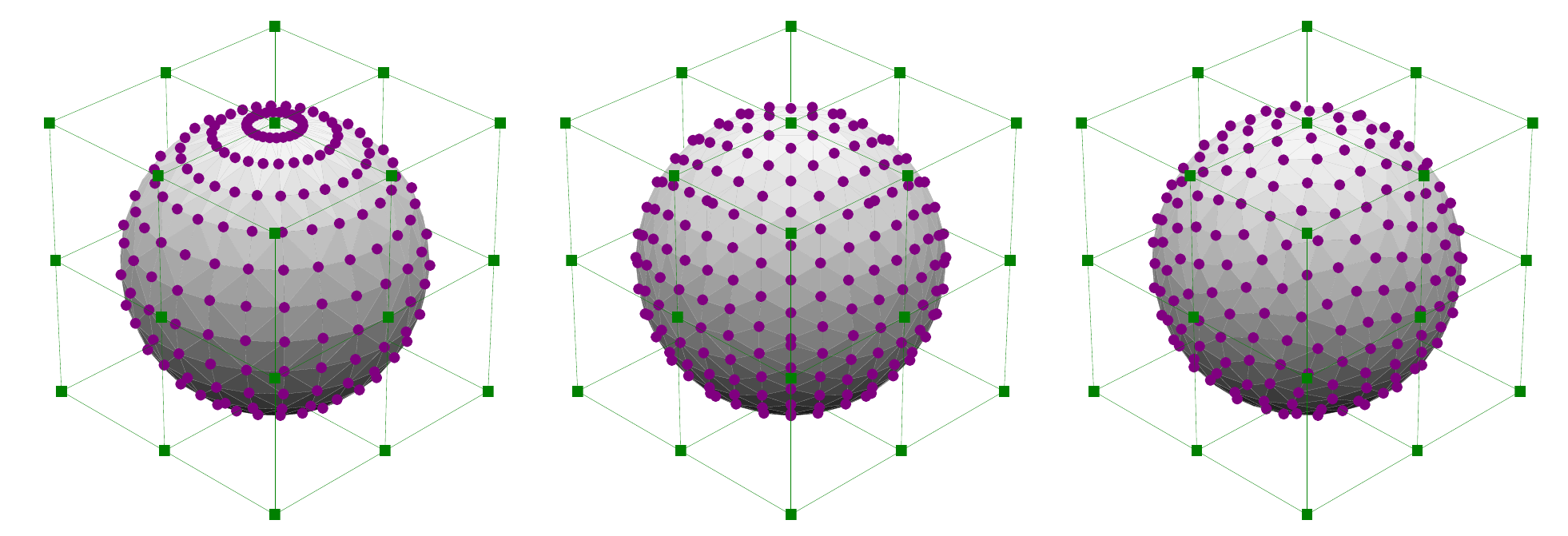}
  \caption{Examples of three-dimensional velocity stencils, comparing the
    distribution of the velocity directions (purple dots) for three
    different quadrature typologies; from left to right we respectively
    show an example for Gauss-Legendre, Lebedev, and spherical design.
    Analogous to Fig. \ref{fig:stencil2D} the purple dots are located on
    a sphere of radius $c\Delta t$ and the green squares denote the cell
    centres. }
  \label{fig:stencil3D}
\end{figure*}

The application of the Gauss-Legendre quadrature to a 2-sphere can be
obtained by making use of the product quadrature rule. Exploiting the
separability of spherical harmonics, the integrals in
Eq. (\ref{eq:sph-quad-cond}) can be expressed as the product of a
circular function $e^{i m \phi}$ and Legendre polynomials:
\begin{align}
  \int_{4\pi} Y_\ell^m(\theta, \phi) d \Omega &\propto
  \left(\int_0^{\pi} P_\ell^m(\cos\theta) \sin\theta d \theta \right) 
  \left( \int_0^{2\pi} e^{i m \phi} d \phi \right) \,,  
\end{align}
where $P_\ell^m(\cos(\theta))$ are the associated Legendre
polynomials. At this point, the two integrals can be evaluated using two
one-dimensional quadrature rules, where the integral in the direction
$\theta$ is performed with a one-dimensional Gauss-Legendre quadrature
\citep{hildebrand-book-1956}, while the integral in $\phi$ can be
evaluated, for example, with the trapezoidal rule. As it is apparent from
the left panel of Fig. \ref{fig:stencil3D}, the Gaussian-Legendre
approach generates an accumulation of points near the north and south
poles of the sphere, and it is less efficient than the other two
quadratures, in the sense that it requires a larger number of points to
achieve the same order of precision.

The Lebedev quadrature (central panel of Fig. \ref{fig:stencil3D}), on
the other hand, follows a different approach: instead of considering the
product of two single quadratures, the integrals in
Eq. (\ref{eq:sph-quad-cond}) are used to build a system of nonlinear
equations in the variables $ \{ w_k, \theta_k, \phi_k \} $. The central
intuition (due to \cite{sobolev-smd-1962}) is that the number of
nonlinear equations can be greatly reduced by considering only the
spherical harmonics of degree $\leq p$ that exhibit invariance under all
transformations that belong to a pre-determined group $G$. This procedure
generates quadratures that are invariant under $G$, that is, these
stencils evaluate exactly both $Q(h)$ and $Q(g(h))$ for all elements $g
\in G$,and still retain the same order of integration $p$. In its
original definition Lebedev's quadrature is by construction invariant
under the octahedral group
\citep{lebedev-ussrcmmp-1975,lebedev-ussrcmmp-1976,lebedev-smj-1977},
although quadratures based on different symmetry groups are also present
in the literature (see, \eg \citep{ahrens-prs-2009} for a quadrature
based on the icosahedral group). From the central panel of
Fig.~\ref{fig:stencil3D} one appreciates that points in Lebedev's
quadratures offer a more homogeneous distribution over the sphere with
respect to the Gauss product rule. Nevertheless, a few points can be seen
to be almost overlapping.

Lastly, we consider the spherical-design quadrature rules, first
introduced by \citet{delsarte-gd-1977}. This type of quadrature requires
that the weights associated to all nodes be equal. The task is then to
define a minimum set of points which integrates correctly all the
spherical harmonics up to order $p$. There is no known rule for the
definition of a generic order $p$ spherical-design quadrature, and the
topic is indeed still object of ongoing research. Nevertheless, numerical
results leading to the definition of quadrature rules up to very high
order are available online. In this work we refer to the set of stencils
presented by \citet{womersley2018}, for which we provide an example in
the right panel of Fig.~\ref{fig:stencil3D}.

\subsection{Numerical procedure}
\label{sec:numerical}

Having presented the equations to be solved and their discretization, we
proceed to summarise the steps required to evolve the LB-discretized RTE.

More specifically, Eq. (\ref{eq:rte_discrete}) can be solved following
the standard stream-and-collide approach, where the streaming step is
performed first, yielding provisional values of the $N_{\mathrm{pop}}$
intensities, \ie
\begin{align}\label{eq:rte-stream-collide}
  I^{*}_i(\bm{r},t) &= I_i(\bm{r}-\bm{\hat{n}}_i \Delta t ,t) \, . 
\end{align}
However, since the discrete velocities fall off-grid for the above
discussed stencils, an interpolation is required. For simplicity we
consider a trilinear interpolation scheme, from which follows
\begin{align}\label{eq:interp}
I_i^*&(\bm{r}-\bm{\hat{n}}_i \Delta t ,t) = \frac{1}{\Delta x\, \Delta y\, \Delta z} \times \Big\{ \Big.                    \nonumber \\
  &I_i(\bm{r}              - \bm{\hat{x}}  - \bm{\hat{y}} - \bm{\hat{z}}, t) \Big(~\,\hphantom{\Delta x - }\Delta t \big| \hat{n}_i^x \big| \Big)
                                                \Big(~\,\hphantom{\Delta y - }\Delta t \big| \hat{n}_i^y \big| \Big)
                                                \Big(~\,\hphantom{\Delta z - }\Delta t \big| \hat{n}_i^z \big| \Big) \nonumber \\
  &I_i(\bm{r}~\,\hphantom{ - \bm{\hat{x}}} - \bm{\hat{y}} - \bm{\hat{z}}, t) \Big(             \Delta x -  \Delta t \big| \hat{n}_i^x \big| \Big)
                                                \Big(~\,\hphantom{\Delta y - }\Delta t \big| \hat{n}_i^y \big| \Big)
                                                \Big(~\,\hphantom{\Delta z - }\Delta t \big| \hat{n}_i^z \big| \Big) \nonumber \\
  &I_i(\bm{r} - \bm{\hat{x}}~\,\hphantom{ - \bm{\hat{y}}} - \bm{\hat{z}}, t) \Big(~\,\hphantom{\Delta x - }\Delta t \big| \hat{n}_i^x \big| \Big)
                                                \Big(             \Delta y -  \Delta t \big| \hat{n}_i^y \big| \Big)
						\Big(~\,\hphantom{\Delta z - }\Delta t \big| \hat{n}_i^z \big| \Big) \nonumber \\
  &I_i(\bm{r} - \bm{\hat{x}}- \bm{\hat{y}} ~\,\hphantom{ - \bm{\hat{z}}}, t) \Big(~\,\hphantom{\Delta x - }\Delta t \big| \hat{n}_i^x \big| \Big)
						\Big(~\,\hphantom{\Delta y - }\Delta t \big| \hat{n}_i^y \big| \Big)
						\Big(             \Delta z -  \Delta t \big| \hat{n}_i^z \big| \Big) \nonumber \\
  &I_i(\bm{r}~\,\hphantom{  - \bm{\hat{x}}- \bm{\hat{y}}} - \bm{\hat{z}}, t) \Big(             \Delta x - \Delta t \big| \hat{n}_i^x \big| \Big)
						\Big(             \Delta y -  \Delta t \big| \hat{n}_i^y \big| \Big)
						\Big(~\,\hphantom{\Delta z - }\Delta t \big| \hat{n}_i^z \big| \Big) \nonumber \\
  &I_i(\bm{r}~\,\hphantom{  - \bm{\hat{x}}}- \bm{\hat{y}}~\,\hphantom{ - \bm{\hat{z}}}, t) \Big(             \Delta x - \Delta t \big| \hat{n}_i^x \big| \Big)
						\Big(~\,\hphantom{\Delta y - }\Delta t \big| \hat{n}_i^y \big| \Big)
						\Big(             \Delta z - \Delta t \big| \hat{n}_i^z \big| \Big) \nonumber \\
  &I_i(\bm{r} - \bm{\hat{x}}~\,\hphantom{ - \bm{\hat{y}} - \bm{\hat{z}}}, t) \Big(~\,\hphantom{\Delta x -}\Delta t \big| \hat{n}_i^x \big| \Big)
						\Big(              \Delta y - \Delta t \big| \hat{n}_i^y \big| \Big)
						\Big(              \Delta z - \Delta t \big| \hat{n}_i^z \big| \Big) \nonumber \\
  &I_i(\bm{r}~\,\hphantom{ - \bm{\hat{x}} - \bm{\hat{y}} - \bm{\hat{z}}}, t) \Big(             \Delta x -  \Delta t \big| \hat{n}_i^x \big| \Big)
                                                \Big(             \Delta y -  \Delta t \big| \hat{n}_i^y \big| \Big)
                                                \Big(             \Delta z -  \Delta t \big| \hat{n}_i^z \big| \Big)
  \Big\} \, ,
\end{align}
with
\begin{align}
	\bm{\hat{x}} &= \big[\, \mathrm{sgn}\big(\hat{n}_i^x\big)\Delta x,\, 0,\, 0\, \big]^T  \\
	\bm{\hat{y}} &= \big[\, 0,\, \mathrm{sgn}\big(\hat{n}_i^y\big)\Delta y,\, 0\, \big]^T  \\
	\bm{\hat{z}} &= \big[\, 0,\, 0,\, \mathrm{sgn}\big(\hat{n}_i^z\big)\Delta z\,\big]^T \, .
\end{align}
For the two dimensional case a bilinear interpolation is used equivalently.

Next, the macroscopic moments are computed according to Eq.
(\ref{eq:moments_discrete}), from which the source term and the
collisional operator are computed using Eq. (\ref{eq:source-and-equil}).

The collision step is then performed as follows:
\begin{align}\label{eq:collision-step}
I_i(\bm{r},t+\Delta t) &= I^{*}_i(\bm{r},t) 
                       - \kappa_0 \Delta t \left[ I^{*}_i(\bm{r},t) - I_i^{\rm eq}(\bm{r},t) \right]    \notag    \\ 
                       &+ \Delta t S_i(\bm{r}, t) \, . 
\end{align}
Note that since in most astrophysical applications the emissivities and
opacities vary over several orders of magnitude, these source terms may
take values much larger than the evolved variable $I_i$, thus making the
RTE a stiff equation that requires special treatment to be solved
efficiently \citep[see, \eg][]{Weih2020b}.

Thus, rather than solving the explicit Eq. (\ref{eq:collision-step}), we solve
the implicit form, \ie
\begin{align}
\label{eq:rte_implicit}
  I_i(\bm{r} + \bm{\hat{n}}_i\Delta t, t + \Delta t) 
  &= 
  I_i^*(\bm{r},t) \notag \\
	&- \kappa_0 \Delta t \left[   I_i(\bm{r},t+\Delta t) 
                                          - I_i^{\rm eq}(\bm{r},t +\Delta t) 
                                   \right] \notag \\
  & + \Delta t S_i(\bm{r},t+\Delta t) \, , 
\end{align}
Because of the presence of the first two moments in the definition of the
source term, we obtain a system of $N_{\mathrm{pop}}$ linear equations
with $N_{\mathrm{pop}}$ unknowns. This could be solved via the inversion
of an $N_{\mathrm{pop}}\times N_{\mathrm{pop}}$-matrix at every
grid-point, which is, however, computationally unfeasible, especially in
3D, when a large number $N_{\mathrm{pop}}$ of discrete velocity
directions is required.
An alternate and computationally much more viable method is known as
\textit{Lambda iteration} \citep{Rampp97}, which works as follows:
\begin{enumerate}
  \item[1.] Take as initial guess for $E(\bm{r},t+\Delta t)$ and
    $\bm{F}(\bm{r},t+\Delta t)$ the moments computed from $I_i$ at time
    $t$.
  \item[2.] Eq. (\ref{eq:rte_implicit}) constitutes a system of
    $N_{\mathrm{pop}}$ linear decoupled equations that can be solved
    analytically and independently, to obtain a first estimate of $I_i$
    at time $t+\Delta t$.
  \item[3.] Use this estimate of $I_i(\bm{r}, t+\Delta t)$ to formulate
    an improved guess for $E(\bm{r},t+\Delta t)$ and
    $\bm{F}(\bm{r},t+\Delta t)$.
  \item[4.] Cycle back to step 2. and repeat until all $I_i(\bm{r},
    t+\Delta t)$ converge, with an error below a prescribed threshold.
\end{enumerate}
To summarise, our scheme consists of two steps; first the streaming-step
according to Eq. (\ref{eq:interp}) and then the collision-step according
to Eq. (\ref{eq:collision-step}), which is solved following the iterative
procedure described above. If the radiation is coupled to a fluid, the
radiative four-force that enters the standard equations of relativistic
(magneto-) hydrodynamics (RMHD) can be computed after every timestep from
the moments $E$ and $\bm{F}$, which themselves are computed approximately
from $I_i$ according to Eq. (\ref{eq:moments_discrete}). At the beginning
of the next timestep, the coefficients $\eta$, $\kappa_a$, $\kappa_0$ and
$\kappa_1$ can then be computed from the updated fluid variables.

As a result, the coupling to a standard RMHD code is exactly\footnote{
Special attention has to be paid to the case of a moving background fluid
(see Sec. \ref{sec:coupled} and appendix \ref{sec:appendix2}).}
the same as
for the commonly used M1 scheme \citep[see][for a detailed description of
  this coupling]{Weih2020b}. Finally, we note that while we restrict
ourselves for simplicity to a first-order time-stepper, it is
conceptually straightforward to extend the evolution to higher orders
using for example implicit-explicit (IMEX) schemes \citep{Pareschi2010},
where the streaming-step is treated explicitly and the collision-step
implicitly.

\section{Numerical tests: static fluid}
\label{sec:tests}
In astrophysical simulations, the ordinary fluid interacting with the
radiation features optical depths that vary considerably, ranging from
the optically thin regime -- where radiation is in free streaming -- to
the optically thick regime -- where radiation is coupled with the fluid
and propagates by diffusion. While in several studies only one of these
regimes is considered, \citep[see \eg][]{Fragile2012,Roedig2012}, it is
our aim to provide a numerical method that can handle both limits, as
well as the intermediate regime. The latter is particularly difficult to be
described accurately by moment-based schemes, mostly because
the closure relation -- which is essential and inevitable in these schemes
-- is normally defined in either the optically thin or the optically
thick regime and is then interpolated between these two limits
\citep{Weih2020b}.

In the series of tests presented below, we discuss the performance of the 
LB method in these different limits, starting in
Sec. \ref{sec:free-streaming} with the optically thin one -- which
represents the most difficult challenge. The optically thick regime is
tested in Sec.~\ref{sec:diffusion_limit} -- where the LB method performs
extremely well. Finally, in Sec.~\ref{sec:rad_sphere}, we present an
example of the intermediate regime, where LB is shown to provide very
accurate results, at variance with the commonly used M1 scheme.

\subsection{Optically thin limit}
\label{sec:free-streaming}

Since the LB method is designed for collisional fluids, we begin the
verification of the LB method and its implementation with a number of
code tests in the most difficult regime, namely, the one in which the
radiation is actually freely streaming, as is the case when
$\eta=\kappa_a=\kappa_0=\kappa_1=0$.

In other words, in a radiative-transport application, the LB method works
best when the emission, absorption and scattering of the radiation with
an underlying fluid is actually taking place, which obviously is not the
case in the free-streaming regime. Nevertheless, since free-streaming is
ultimately taking place in any astrophysical scenario of interest, such
as supernova explosions and neutron-star mergers -- where the radiation
is composed of neutrinos -- it is clear that testing the LB method in
this regime is most important.

\subsubsection{Beam tests}
\label{sec:beam_tests}

\begin{figure}
  \includegraphics[width=1.0\columnwidth]{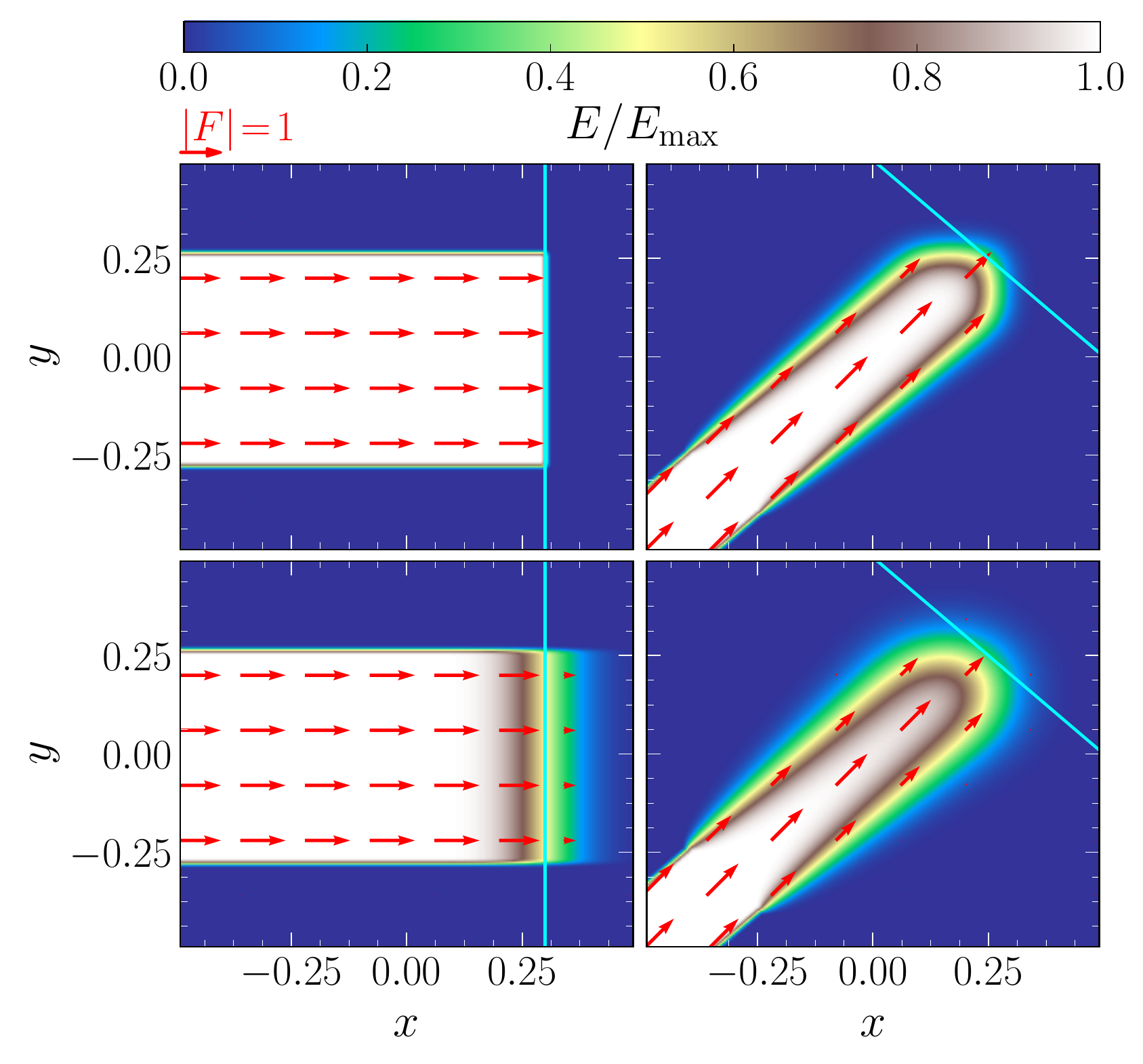}
  \caption{\textit{Left:} Straight beam of radiation with a CFL-number of
    $1.0$ (top) and a CFL-number of $0.2$ (bottom). Colour coded is the
    energy-density, while the momentum density is shown by red
    arrows. The cyan line indicates, how far the beam should have
    propagated until $t=0.7$. \textit{Right:} The same as the left panel,
    but for a beam propagating diagonally. The region with $x,y<-0.25$ is
    frozen via a boundary condition in order to continuously shoot the
    beam into the grid from the bottom left.}
  \label{fig:beams}
\end{figure}

\begin{figure}
  \includegraphics[width=1.0\columnwidth]{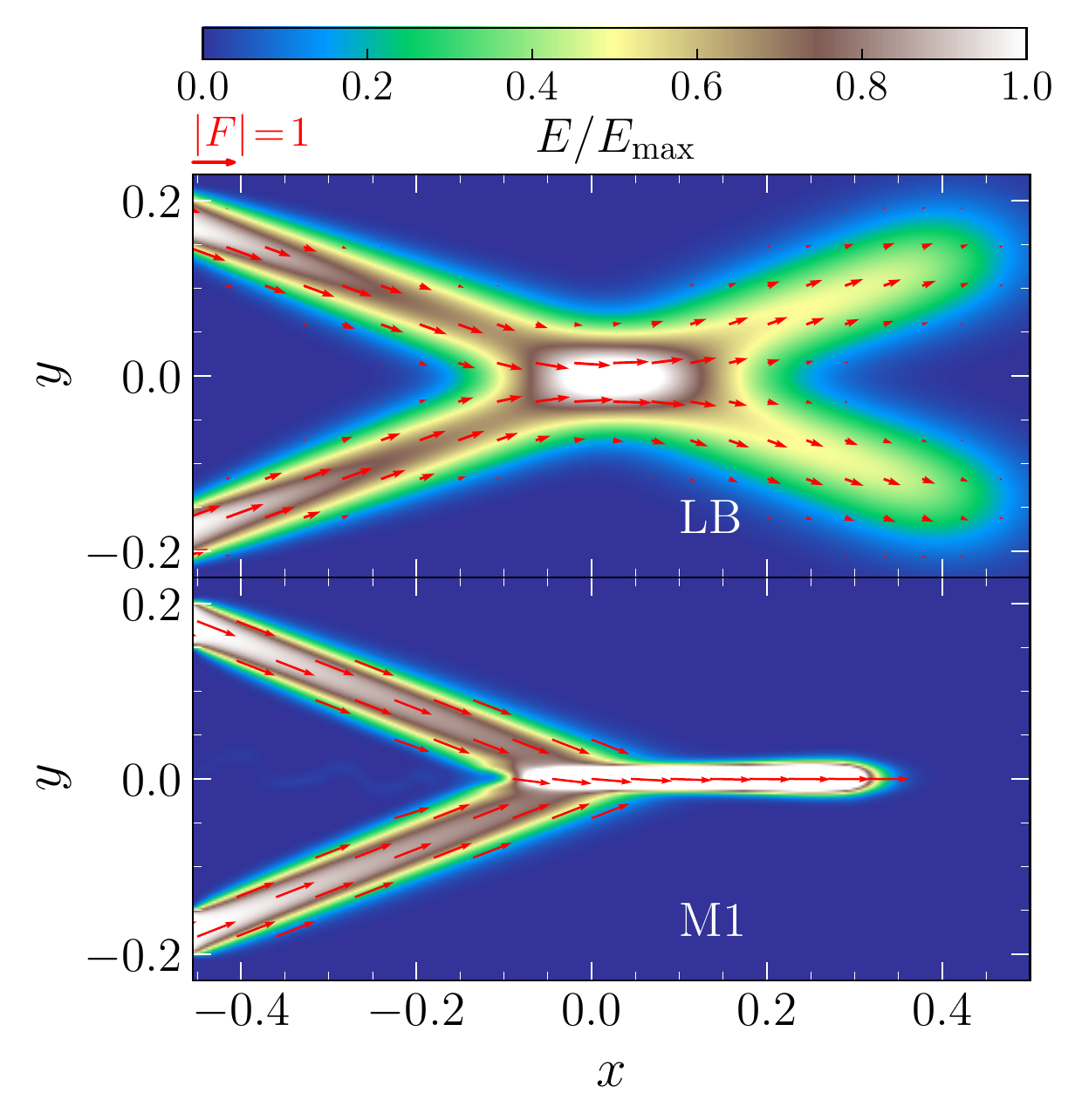}
  \caption{Comparison of the performance for the beam-crossing problem
    between LB (top) and M1 (bottom). }
  \label{fig:crossing_beams}
\end{figure}

We start with a ``classical'' beam test, namely, the propagation of a
well defined beam of radiation injected from the boundary of the
computational domain. Having set $\eta=\kappa_a=\kappa_0=\kappa_1=0$,
Eq. (\ref{eq:rte_implicit}) clearly states that free-streaming is
trivially achieved when each of the intensities $I_i$ is propagated from
one grid-cell to the next following the underlying stencil. The top-left
panel of Fig.~\ref{fig:beams} shows an example of a freely streaming beam
on a grid of size $-0.5<x<0.5,\, -0.5<y<0.5$, which we cover with
$100^{2}$ equal-size grid cells. This test is performed in 2D using the
stencil shown in Fig.~\ref{fig:stencil2D} with $N_{\mathrm{pop}}=8$
discrete velocity directions. Throughout the simulation we constantly
inject a radiation beam from the grid's left boundary. To do so, at all
times we enforce the following condition:
\begin{align}
  I_i =
  \begin{cases}
    1 \quad &  i=0 \\
    0 \quad &  i\neq 0 \,, 
  \end{cases}
\end{align}
at $x=-0.5$ and $|y|<0.25$.

As expected, the beam propagates parallel to the $x$-axis from left to
right at the speed of light. In Fig. \ref{fig:beams} we show with a
colorcode the radiation energy density and with arrows the momentum
density, as computed according to Eq. (\ref{eq:moments_discrete}) at time
$t=0.7$, and where we have used a timestep $\Delta t = \Delta x = \Delta
y$. This timestep is a standard choice in classical LB methods and leads
to perfect streaming, \ie no diffusion along the beam's direction of
propagation. This is due to the fact that the streaming step, \ie
Eqs. (\ref{eq:rte-stream-collide}) and (\ref{eq:interp}), reduces to
\begin{equation}
  I_i^*(\bm{r}, t+\Delta t) = I_i(\bm{r}-\hat{\bm{n}}_i\Delta t, t) = 
  \frac{|\hat{n}_i^x|}{\Delta x} \times I_i(\bm{r}-\bm{\hat{x}},t) \, .
\end{equation}
Considering that $|\hat{n}_i^x| / \Delta x = 1$, where $|\hat{n}_i^x|$ is
the $x$-th component of the $i$-th velocity vector, we find that the
intensity is simply propagated from one cell to its right neighbour
during each iteration.

Most of the astrophysical codes solving the equations of RMHD employ
either finite-volume or finite-differencing schemes and require a
timestep $\Delta t=\mathrm{CFL}\, \Delta x$ for numerical stability,
where $\rm CFL \leq 1$ is the Courant-Friedrichs-Lewy coefficient
\citep{Rezzolla_book:2013}. Keeping in mind that for simulations of
astrophysical systems, the LB method needs to be coupled to such a RMHD
code, we also perform the above beam tests with $\Delta t = 0.2 \Delta
x$, where $\mathrm{CFL}=0.2$ is a typical value chosen in
multidimensional relativistic hydrodynamics. This amounts to squeezing
the stencil and requires the complete evaluation of Eq. (\ref{eq:interp})
during the streaming step.

The result of this simulation is reported in the bottom left panel of
Fig. \ref{fig:beams}. In contrast to the case of perfect streaming
reported in the top panel, the bottom panel shows that there is some
diffusion of radiation ahead of the beam 
(see, for comparison, Fig. 1 of \cite{Weih2020b} for the same
behaviour in an M1 code). Note that even though the previous test is a
physically trivial one, it is nonetheless very useful to verify the correct 
implementation of the streaming step and the interpolation, according to
Eq. (\ref{eq:interp}).

A more challenging setup is that of a beam propagating along a direction
not parallel to any coordinate axis. To model this case, we choose the
same setup as above, but enforcing $I_1$, which corresponds to
$\bm{\hat{n}}_1=[\mathrm{cos}(\pi/4),\,\mathrm{sin}(\pi/4)]$, rather than
$I_0$ to a nonzero value; of course $I_{i \neq 1}=0$.

Results are presented in the right panels of Fig. \ref{fig:beams}, where
we show again the cases $\Delta t=\Delta x$ (top panel) and $\Delta t =
0.2\Delta x$ (bottom panel). In both cases, the beam now diffuses much
more, an effect that can also be observed for the commonly used
moment-schemes \citep{Weih2020b}.
It is worth remarking that in all of the above tests, the beams propagate
along one of the discrete velocity directions. However, radiation beams
can propagate in any direction in the continuum, not necessarily along a
discretized velocity direction. In this case, the LB scheme would
inevitably incur increasing errors as we approach the optically thin
regime.
These can be tamed by developing high-order phase-space interpolators,
possibly involving non-local neighbours both in configuration and velocity
space. Clearly, this exposes a tension between accuracy and efficiency
which still needs to be explored and resolved in full. In this paper, we
rely upon trilinear and nearest-neighbour interpolation in configuration
and velocity space, respectively.

Next, we consider the performance of our code for another ``classical''
and yet fundamental free-streaming test: two crossing beams. Assuming
the radiation to consist of photons or neutrinos of the same flavour, one
would expect the two beams to cross each other without interacting. The
M1 scheme is known to perform very poorly under these conditions
\citep{Fragile2014,McKinney2014,Foucart2015a,RiveraPaleo2019,Weih2020b},
due to the fact that it retains only the two lowest moments, hence only
the average direction of propagation.

This information could in principle be obtained by employing higher
moments, at the cost of increased computational costs. On the other hand,
in an LB method the various directions of propagation are evolved
separately and thus crossing beams can be correctly evolved in time. This
is shown in Fig. \ref{fig:crossing_beams}, where the results from an LB
simulation (top panel) are compared to the ones from an M1 code (bottom
panel). In both cases, we perform the simulation on a grid of size
$-0.5<x<0.5$ and $-0.25<y<0.25$ with a resolution of $200\times 100$ and
choose again a CFL coefficient of $0.2$. For the M1 solution, we
initialise the momentum densities of the two beams to point towards the
center, while for LB we simply initialise the intensities for the
corresponding directions.

It can be seen clearly from Fig. \ref{fig:crossing_beams} that in the LB
simulation the two radiation beams cross as expected, while they merge to
an averaged beam when using M1. The proper description of this behaviour
is of crucial importance in astrophysical simulations, where beams of
radiation -- such as those emitted from the torus of a binary
neutron-star merger remnant or in a supernova explosion -- are expected
to meet and interact. Hence, the successful outcome of this test provides
encouraging evidence that the present LB method can be used in the place
of moment-based schemes.

\subsubsection{Radiation wave in free-streaming regime}
\label{sec:radiation_wave}

\begin{figure*}
  \includegraphics[width=0.99\textwidth]{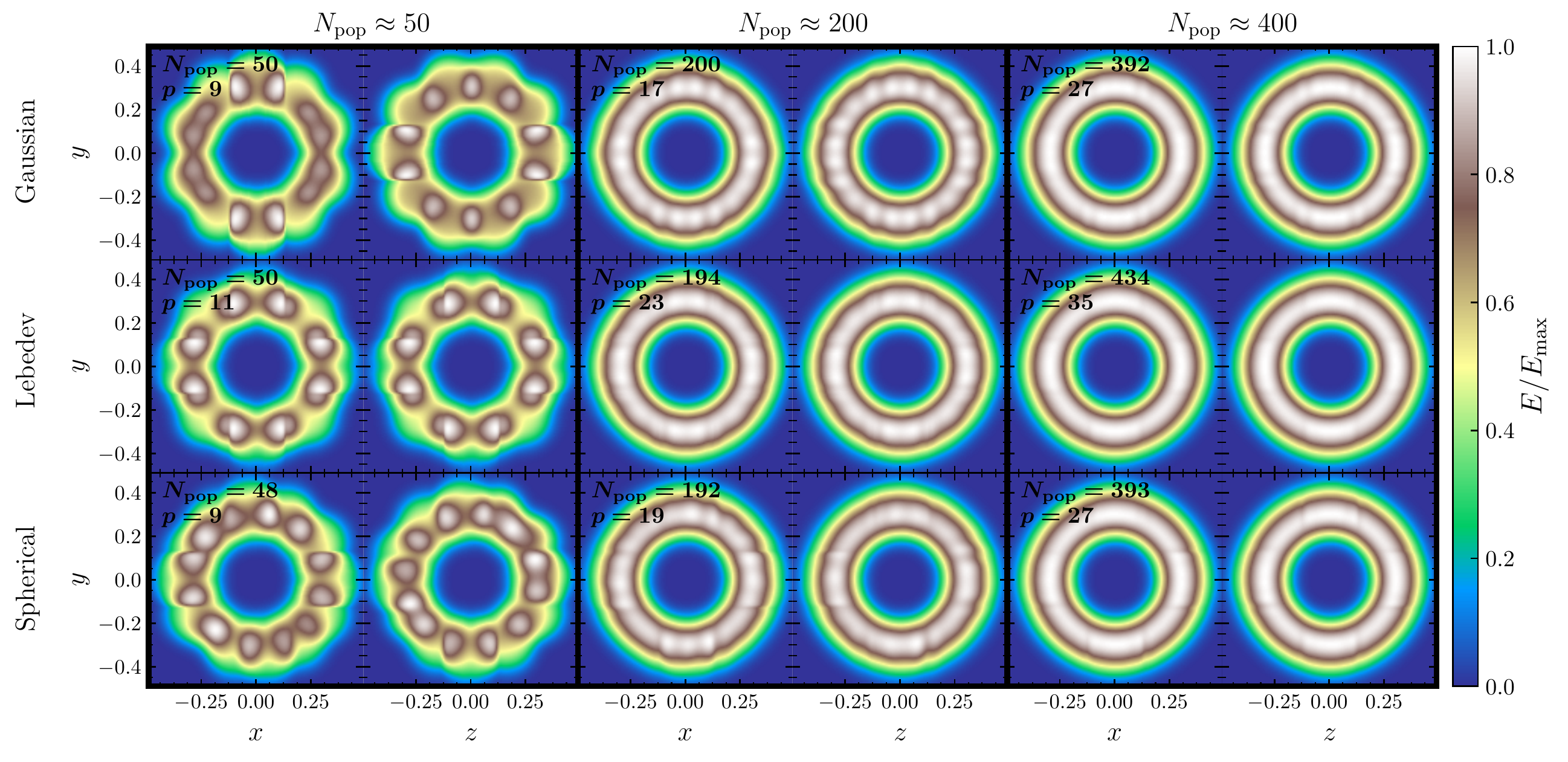}
  \caption{Spherical freely-streaming wave for different types of
    stencils. From top to bottom we show stencils derived from (i)
    Gaussian product quadrature, (ii) Lebedev quadrature and (iii)
    spherical design. Each column compares these three stencils for a
    comparable number of discrete velocities. Snapshots show the
    radiation energy-density in the $(x,y)$ (left) and $(y,z)$ (right)
    plane after 200 iterations. In every panel we also report
    $N_{\mathrm{pop}}$ and the quadrature order $p$. }
  \label{fig:free_wave}
\end{figure*}
In order to evaluate if radiation is propagated isotropically by the
numerical scheme, we show here the results obtained for the case of a
spherically symmetric propagation of a radiation wave. The wave is
expected to expand at the speed of light in all directions, with its
energy density decreasing over time following an inverse-square law for
the distance travelled by the wavefront.
We run all simulations on a 3D grid with $200^3$ uniform sized grid cells
with $\Delta t = 0.2$, and initialise the intensities by setting $I_i=1$
within a sphere of radius $R=16\Delta x$ grid-cells and $I_i=0$
everywhere else. Stencils based on the three quadrature methods
introduced in Sec.~\ref{sec:methods2} are employed, with different
quadrature orders, so as to compare both the methods and the various
orders.

Figure \ref{fig:free_wave} shows the results of these simulations in
terms of the radiation energy-density, after $200$ iterations, in the
$(x,y)$ and $(y,z)$ planes. It is clear that a low number of discrete
velocities $N_{\mathrm{pop}}$ in the stencils does not allow the
radiation to stream isotropically. However, upon increasing
$N_{\mathrm{pop}}$, the isotropy of the system rapidly improves up to
very satisfactory levels. Also to be noted, none of the three methods
emerges as a neat winner, all yielding results of similar quality. This
is true even for the Lebedev quadrature, which features a consistently
higher quadrature order as the other two types for the same
$N_{\mathrm{pop}}$.

\begin{figure}
  \includegraphics[width=0.99\columnwidth]{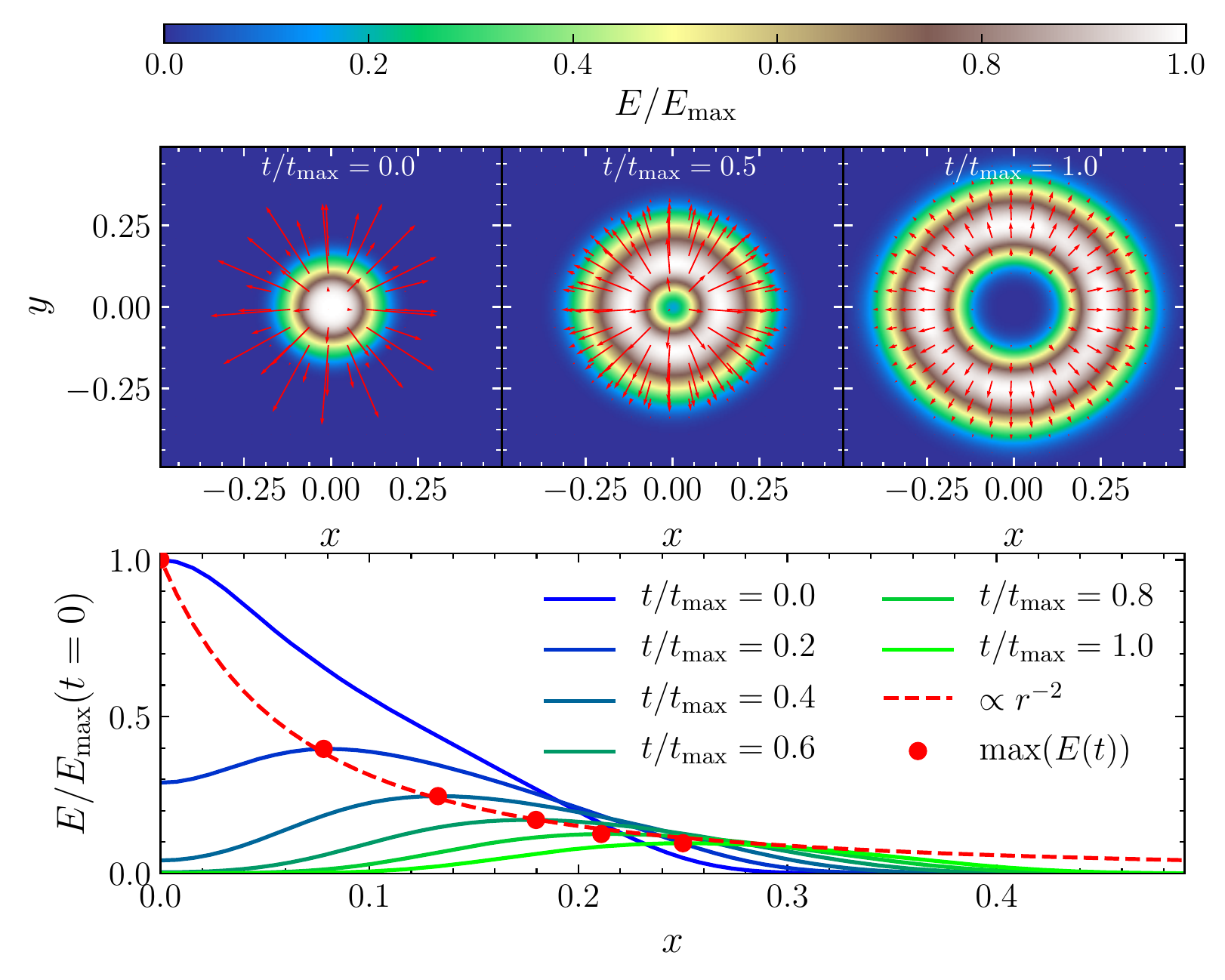}
  \caption{\textit{Top:} Radiation energy density (colour coded) and
    momentum density (red arrows) at three representative times in the
    (x,y) plane for the radiation wave test with the Lebedev quadrature
    of order $p=23$. \textit{Bottom:} Diagonal profiles of the energy
    density at different times (blue to green) and inverse square law
    (red-dashed) fitted to the maxima of each time (red dots).}
  \label{fig:free_wave2}
\end{figure}
Finally, we show that the wave's maximum energy decays proportionally to
$r^{-2}$, which is expected for this 3D test, where the energy is
conserved and spreads over a spherical shell of area $ \pi r^2$ in
time. Figure \ref{fig:free_wave2} shows this behaviour for the reference
case of the Lebedev quadrature of order $p=23$ (middle panel in
Fig. \ref{fig:free_wave}). The figure reports the profile of the
radiation energy density along a diagonal cut at different times. Note
that the maxima of these cuts align well with an inverse square law
(red-dashed line).

\subsection{Optically thick limit}
\label{sec:diffusion_limit}

Having tested the LB method for the solution of the RTE in the optically
thin regime, we now focus on the optically thick regime. We recall that
an optically thick medium either absorbs (in the case of a high value of
the absorption opacity $\kappa_a$) or scatters radiation (in the case of
a high scattering opacities $\kappa_0$ and $\kappa_1$). As a result, we
present tests of the absorption scenario in Secs. \ref{sec:shadow} and
\ref{sec:rad_sphere} and the scattering scenario, \ie the diffusion
limit, in Sec. \ref{sec:scattering}.

\subsubsection{Shadow test}
\label{sec:shadow}

We start again with a beam using a similar setup as for the straight beam
in Sec. \ref{sec:beam_tests}, but we place an optically thick obstacle on
the beam path. ``Thick obstacle'' means that we set the absorption
opacity to a large value in the region occupied by the obstacle. From a
physical point of view, the obstacle can then be viewed as a black-body
absorbing the incoming radiation without re-emitting it ($\eta=0$).

\begin{figure}
  \includegraphics[width=0.99\columnwidth]{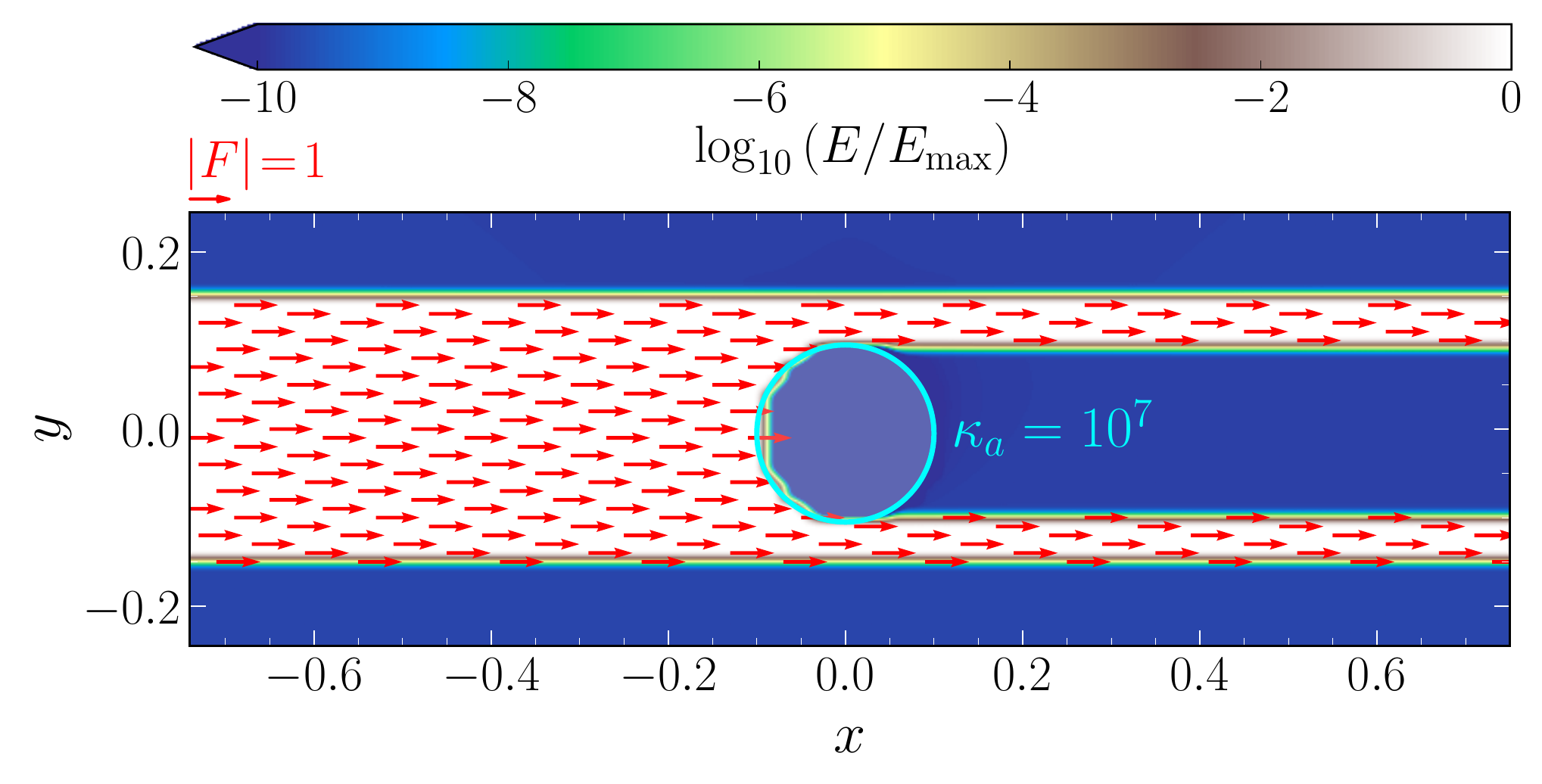}
  \caption{Radiation beam hitting an optically thick obstacle (cyan
    circle) with $\kappa_a=10^7$. The colour encodes the radiation
    energy-density and the red arrows show magnitude and direction of the
    energy momentum-density.}
  \label{fig:shadow}
\end{figure}

More specifically, on a 3D grid of size $-0.75 < x < 0.75,\, -0.25<y,\,
z<0.25$ and covered with $150\times50\times50$ uniform sized grid cells,
we set $\kappa_a=10^7$ within a sphere of radius $R=0.1$, centred at the
origin. The beam is again injected into the grid from the left
boundary. To this end, the 3D stencil must be chosen so as to provide a
discrete velocity direction parallel to the $x$-axis. All other
directions do not matter, since the radiation only propagates parallel to
the $x$-axis in this test. Figure \ref{fig:shadow} shows the beam in the
$(x,y)$ plane after it has propagated across the whole grid and passed
the obstacle marked as a cyan-coloured circle.

It can be seen that, as expected, the beam is blocked by the obstacle and
only a negligible amount of radiation diffuses inside the optically thick
region. This is due to the finite resolution of the grid and is
progressively suppressed as the grid is refined.

We should note that, in contrast with all the tests performed so far, the
set of equations that we solve here is very stiff because of the high
numerical value of $\kappa_a$ (as compared to the evolved variables).
Obtaining a numerically stable solution with a reasonable timestep (we
here use again $\Delta t=0.2\,\Delta x$) is then only possible when using
an implicit solver in time. As described in Sec. \ref{sec:numerical}, we
have implemented the \textit{Lambda iteration}-method. It is found that,
at least for this simple test, it converges in at most three cycles at
every timestep, within a tolerance of $\Delta E/E = 10^{-14}$.

\subsubsection{Radiating sphere}
\label{sec:rad_sphere}
%
\begin{figure*}
  \includegraphics[width=0.9\textwidth]{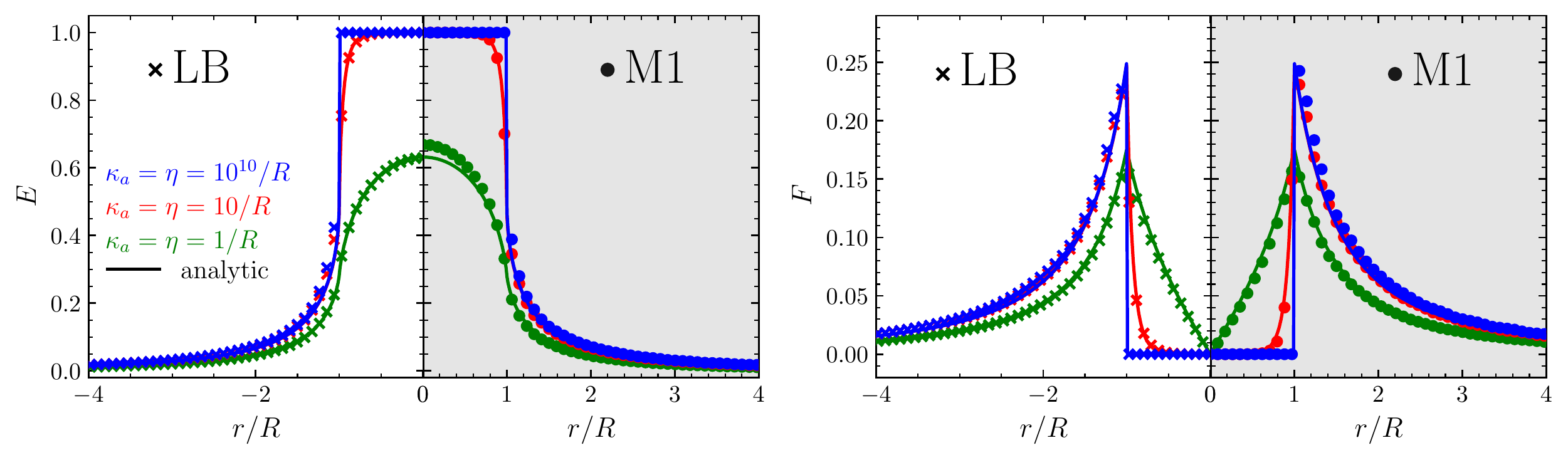}
  \caption{Diagonal cuts for the radiation energy-density (left) and the
    magnitude of the momentum-density (right) of the equilibrium for the
    radiating-sphere test. We compare the numerical solution obtained
    with LB (crosses; left halves) and M1 (circles; right halves) with
    the analytic solution (solid lines.) for a low (green), moderate
    (red) and high (blue) value of $\kappa_a=\eta$. }
  \label{fig:emitting_sphere_profile}
\end{figure*}

We next consider a test involving the emission of radiation as it occurs
for a background static fluid in thermodynamic equilibrium with the
radiation, that is, when the emissivity $\eta$ is equal to the absorption
opacity $\kappa_a$. More specifically, we simulate a system in which
$\eta=\kappa_a=\text{const.}$ within a sphere of radius $R$ and zero
outside the sphere. This is known as the
homogeneous-/radiative-/emitting-sphere test, as first proposed by
\citet{Smit:97}. From a physical point of view, this system can be
thought of as a dense sphere with a sharp boundary to vacuum, as is the
case of a neutron star, that constantly emits radiation from its surface
to the surrounding vacuum.

The system eventually finds a steady state, for which the analytical
solution of the distribution function in terms of radial distance $r$ and
azimuthal angle $\theta$ reads as follows:
\begin{equation}
  f(r,\mu) = b(1-e^{-\kappa_a\, s(r,\mu)}) \, ,
\end{equation}
where $\mu\coloneqq \mathrm{cos}\,\theta$, $b=\kappa_a/\eta=1$ in our
case, and
\begin{equation}
  s:=\begin{cases}
  r\mu + Rg(r,\mu) & r<R ~ \mathrm{and} ~ -1<\mu<1 \,, \\
  2Rg(r,\mu) & r\geq R   ~ \mathrm{and} ~ \sqrt{1-R^2/r^2} <
  \mu < 1 \,,
  \end{cases}
\end{equation}
with
\begin{equation}
  g(r,\mu) := \sqrt{ 1-\frac{r^2}{R^2}(1-\mu^2)}\,.
\end{equation}

Since the final equilibrium is spherically symmetric, the zeroth moment
$E$ can be obtained via the integration of the distribution function over
$\mu$, \ie
\begin{equation}
  E(r) = \frac{1}{2} \int_{-1}^1 d\mu\, f(r,\mu) \,.
        \label{eq:radsph_1}
\end{equation}
Likewise, we obtain the first moment $F$ as
\begin{equation}
  F(r) = \frac{1}{2} \int_{-1}^1 d\mu\, \mu f(r,\mu) \,.
        \label{eq:radsph_2}
\end{equation}

We run this test using a 3D grid
\footnote{As already remarked by \citet{Radice2013} and
  \citet{Weih2020b}, this test requires three dimensions on a Cartesian
  grid, since fluxes also propagate across grid-cells in the angular
  directions.}
with $128^3$ uniform sized grid cells and a spherical-design stencil of
order 20 ($N_{\mathrm{pop}}=222)$ for three different values of
$\kappa_a$, \ie $\kappa_a = R^{-1},\, 10 R^{-1},\, 10^{10} R^{-1}$, and
where $R=1/(8 n_x)$ is the sphere's radius.

The discrete intensities are initialised as
\begin{equation}
  I_i(r,t=0):=  w_i
                \begin{cases}
                  1 		& r<R \,, \\
                  r^{-2} 	& r\geq R \, ,
                \end{cases}
\end{equation}
where $w_i$ are the weights associated with the underlying velocity stencil.

The results of these simulations are reported in
Fig. \ref{fig:emitting_sphere_profile} in terms of the radiation
energy-density (left panel) and of the momentum-density (right panel).

As one can appreciate, the LB method works very well in all of the three
cases. As already discussed for the previous shadow test, numerical
stability in the case of $\kappa=10^{10}/R$ is only possible thanks to the
implicit time-stepper.

Figure \ref{fig:emitting_sphere_profile} also offers a comparison with
the corresponding results obtained with an M1 scheme (these are shown in
the right portions of the two panels in
Fig. \ref{fig:emitting_sphere_profile}). It is clear that for small
values of $\kappa_a$, the LB method performs significantly better than
M1. As already reported by \citet{Weih2020b}, the M1 code fails in this
specific case due to the lack of the second moment, \ie the correct
pressure tensor. While the pressure tensor is exact in the limit of
infinite optical depth (see red and blue curves in
Fig. \ref{fig:emitting_sphere_profile}), it is not so for the intermediate
regime between optically thick and thin media. Indeed, the case of
$\kappa_a=R^{-1}$ (green curves) falls exactly in this regime, for which
the pressure tensor is interpolated inaccurately (see also
\citet{Murchikova2017} for a detailed analysis of this test for the M1
scheme with different closures). The failure of the M1 scheme in this
test is particularly evident upon looking at the energy density, which is
systematically above the correct analytical solution inside the sphere
(note that the tail, which is within the free-streaming regime is
reproduced correctly). The same -- albeit less visible -- is true for the
corresponding momentum density, which is everywhere below the analytic
solution.

\subsubsection{Radiation wave in scattering regime}
\label{sec:scattering}

As a final test of the solution of the RTE with the LB method, we
consider the effects of scattering by simulating a Gaussian distribution
that diffuses over a static background fluid. The same test has been used
also in the validation of various M1 codes \citep[see, \eg][]{Pons2000,
  Oconnor2015, Weih2020b}.
The initial conditions are given by:
\begin{equation}
  E(r, t = 0) = A \exp{ \left( - \frac{ (r - r_0)^2 }{2 \sigma_0^2} \right) } \,,
\end{equation}
where $A$ is the energy-density amplitude and $\sigma_0$ the width of the
Gaussian at $t = 0$ centered at $r_0$. By neglecting emission and
absorption ($\eta = \kappa_a = 0$), the evolution of the system is
governed by the diffusive equation, which exhibits the following analytic
solution:
\begin{equation}\label{eq:gauss-hill-analytic}
  E(r,t) 
  = 
  A \frac{\sigma_0^2}{\sigma_0^2 + \sigma_D^2} 
  \exp{ \left( - \frac{ (r - r_0)^2 }{2 (\sigma_0^2 + \sigma_D^2)} \right) } \,,
\end{equation}
where $\sigma_D = \sqrt{2 D t}$. The connection between the microscopic
parameters $\kappa_0$ and $\kappa_1$ in Eq. (\ref{eq:rte_gray}) and the
diffusion coefficient $D$ is discussed in Appendix~\ref{sec:appendix1} by
an asymptotic analysis.

\begin{figure}[h!]
  \includegraphics[width=0.99\columnwidth]{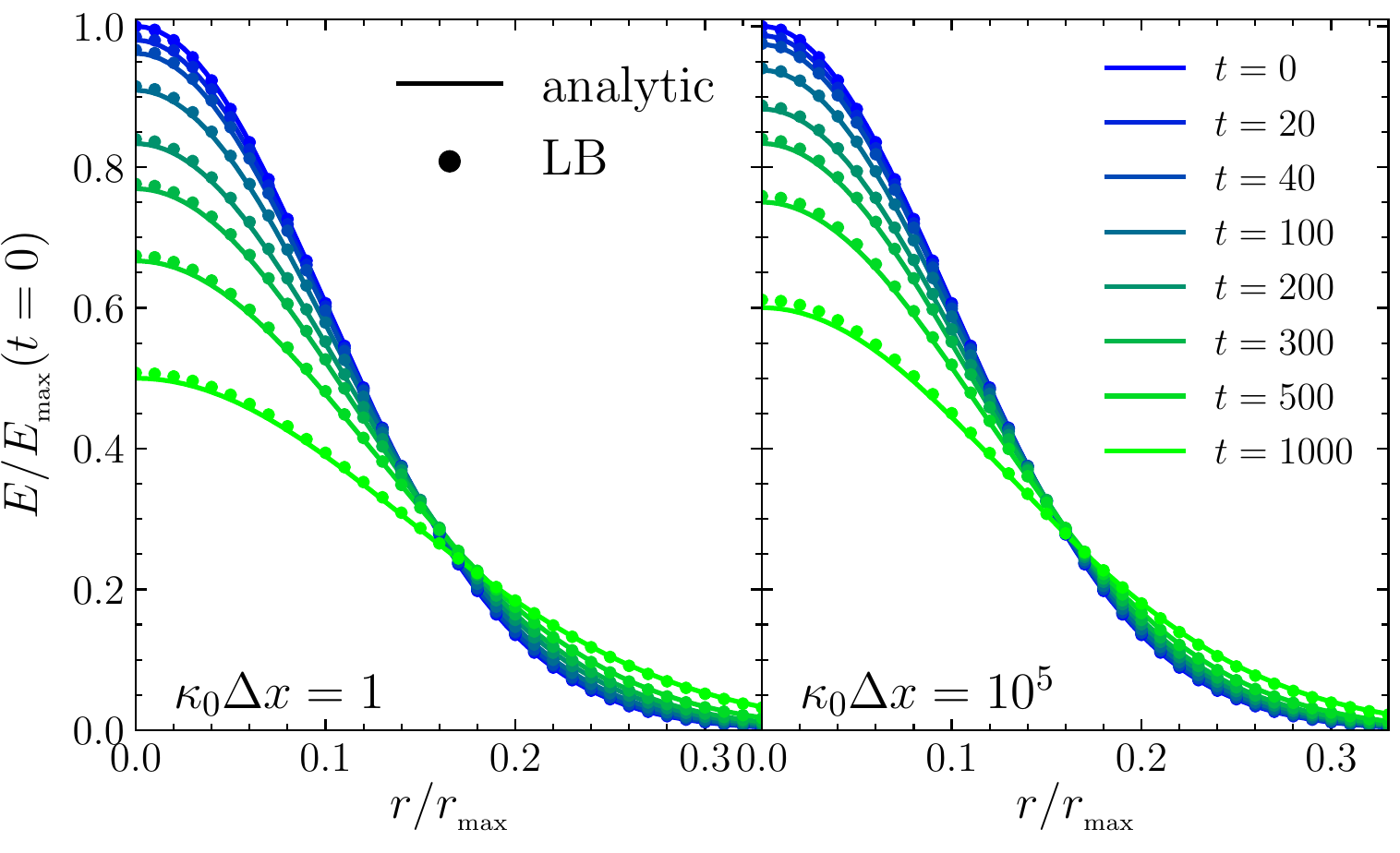}
  \caption{Gaussian wave packet diffusing in a static fluid. The analytic
    solution (solid lines) is compared with results of simulations (dots)
    obtained on a two-dimensional grid of size $100^2$, with $\Delta t =
    0.1$, and $\lambda = 0.5$. The two panels show results at low and
    high Peclet numbers. }
  \label{fig:scattering-wave-2D}
\end{figure}

\begin{figure*}
  \includegraphics[width=0.99\textwidth]{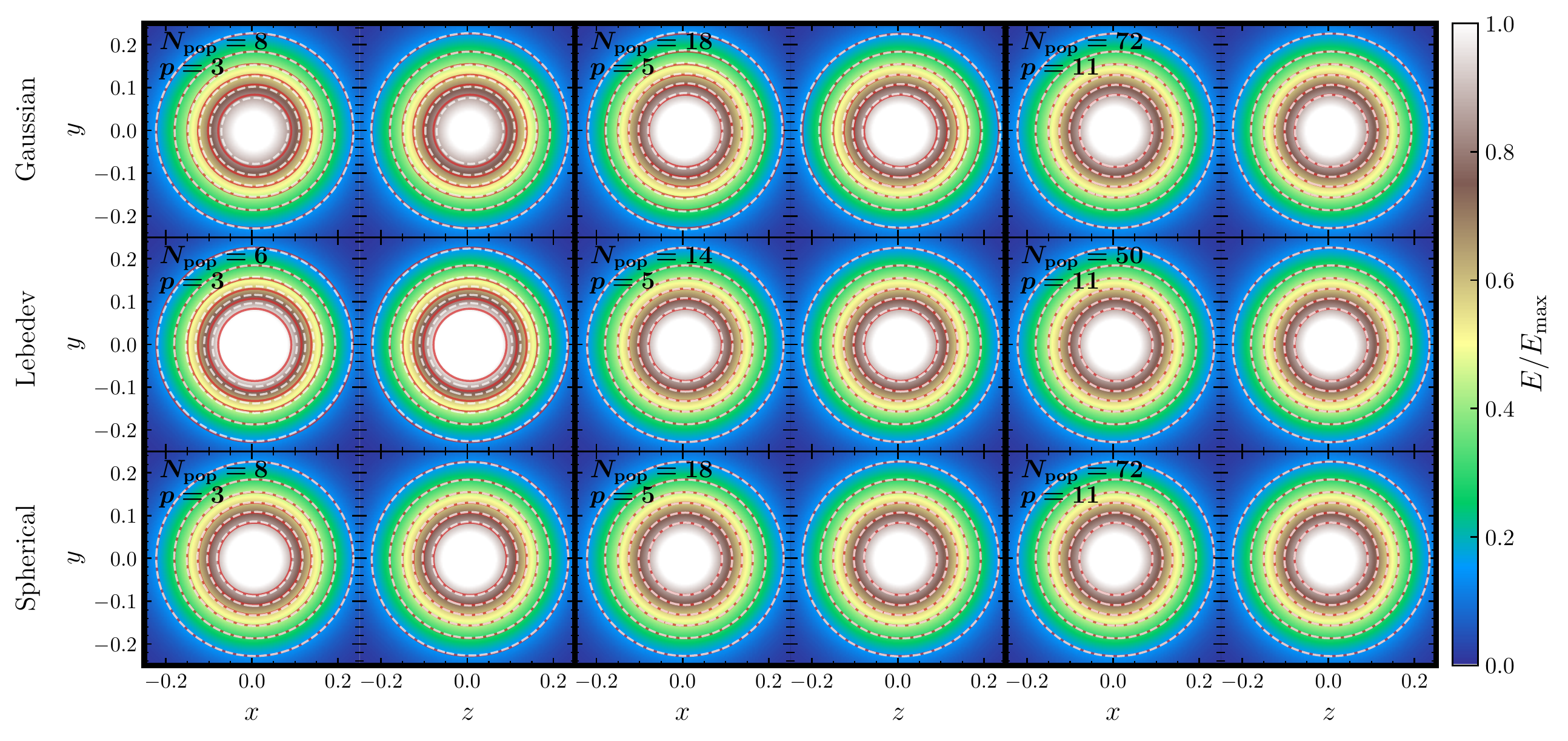}    
  \caption{Scattering wave in 3D. The simulations are performed on a grid
    of size $128^3$, with $\Delta t = 0.1$, $\kappa_0 \Delta x = 1 $ and
    $\lambda = -0.25$. From top to bottom we show the results obtained
    using stencils derived from (i) Gaussian product quadrature, (ii)
    Lebedev quadrature and (iii) spherical design. Each column compares
    these three stencils for a comparable number of discrete
    velocities. Snapshots show the radiation energy-density in the
    $(x,y)$ (left) and $(y,z)$ (right) plane after $100 / \Delta t$
    iterations. Contour lines are used to compare the analytical solution
    (continuous red lines) with the numerical results (white dotted lines). 
    In each panel, we also report $N_{\mathrm{pop}}$ and the quadrature order $p$. 
    }
  \label{fig:scattering-wave-3D}
\end{figure*}

We perform two different simulations at different values of the ``Peclet
number'' ($ \text{Pe} := \kappa_0 \Delta x $), tracking the history of
the system at different time intervals. In
Fig.~\ref{fig:scattering-wave-2D}, we compare the analytical solution
(lines) with the numerical results (dots) obtained running in 2D, on a
$100^2$ grid, with $\kappa_0 \Delta x = 1$ (left panel) and $\kappa_0
\Delta x = 10^5$ (right panel). In both cases, we set $\lambda \coloneqq
3\kappa_1/\kappa_0 = 0.5$.
The same test can be performed in three dimensions, leading to results of
similarly good quality. Finally, in Fig.~\ref{fig:scattering-wave-3D} we
show that, in stark contrast with the results shown for the
free-streaming regime (Fig.~\ref{fig:free_wave}), the evolution is well
captured even by quadratures of comparatively low orders, featuring a
small number of discrete directions. This highlights the fact that the LB
method performs extremely well in the diffusion limit, in fact the one it
was born for. In general, we have observed that in this regime
quadratures with order $p=5$ are sufficient to correctly recover the
correct diffusive dynamics.

\section{Numerical tests: dynamical fluid}
\label{sec:coupled}

\begin{figure}
  \includegraphics[width=0.99\columnwidth]{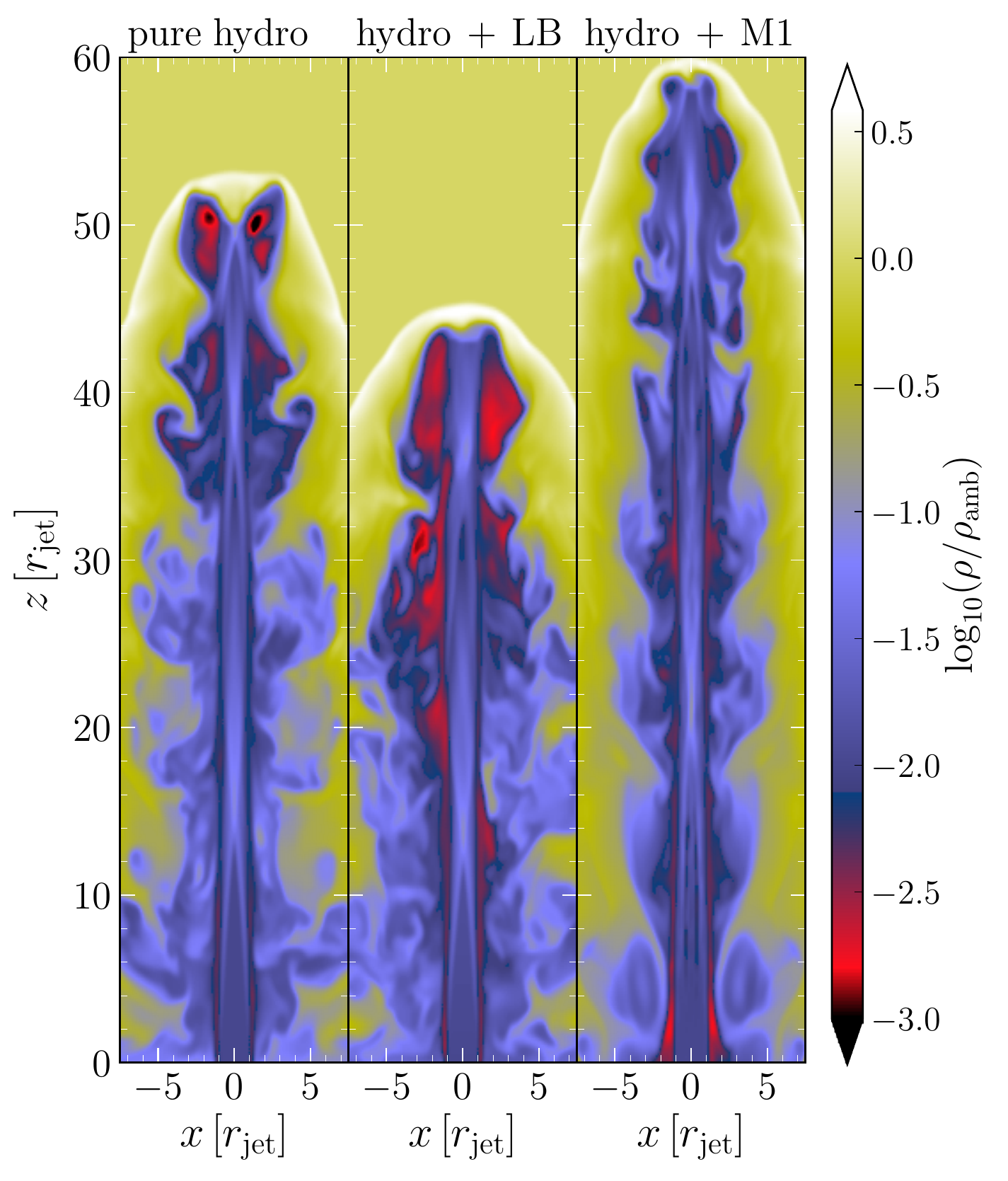}
  \caption{Cut through the $(x,z)$ plane for the relativistic jet after
    $t=125\,r_{\rm jet}$. Shown is the rest-mass density for the
    pure-hydro (left) and the coupled hydro-radiation using LB (middle)
    and M1 (right).}
  \label{fig:jets}
\end{figure}

After having shown that the proposed radiative LB method performs well in
all the regimes of the solution of the RTE on a static fluid, we next
move on to test the method for a dynamical fluid, i.e. one coupled to
radiation via an RMHD simulation of a realistic astrophysical scenario.
As a result, the test presented here is of great importance as it allows
us to explore our novel approach under conditions typical of the
astrophysical scenarios for which it has been developed in the first
place.

We should also note that, strictly speaking, this is not a genuine test,
as the explored scenario does not have an analytic solution to be
compared with. Moreover, simulations of this type have been performed
before only by \citet{RiveraPaleo2019} using an M1 scheme for evolving
the radiation but with rather different prescriptions for the properties
of the radiation field. 

Hence, to gain confidence on the reliability of our results and contrast
them with those obtained when the RTE is solved using an M1 scheme, we
carry out additional simulations of the identical physical scenario but
making use of the M1 code \texttt{FRAC}
\citep{Weih2020b}. While this does not prove the correctness of our
results -- both codes could be incorrect despite the many tests passed --
it does provide us with the confidence necessary to implement the LB
method for even more realistic astrophysical scenarios.

More specifically, we simulate the evolution of a relativistic jet as it
propagates through the interstellar medium. Simulations of this type have
a long history since such jets are of major importance in the study of
active galactic nuclei \cite[see][for a recent review]{Perucho2019},
where they are produced through the accretion process onto a supermassive
black hole \cite[see][for a recent comparative study]{Porth2019_etal}.

Highly energetic relativistic jets are known to accompany the
phenomenology of short gamma-ray bursts and are associated with the
merger of two magnetised neutron stars \citep{Rezzolla:2011}.

Here, we simulate this problem by coupling our LB code to the
\textit{Black hole accretion code} (\texttt{BHAC})
\citep{Porth2017}. \texttt{BHAC} is a finite-volume code that solves the
equations of general-relativistic MHD in a fixed and curved
spacetime. The results presented here, however, are restricted to a flat
background spacetime.

In essence, we perform the coupling between \texttt{BHAC} and the LB code
following the strategy indicated below.

\begin{itemize}
\item[1.] At every iteration, we pass the fluid rest-mass density $\rho$,
  temperature $T$ and three-velocity $\varv_i$ to the LB code, from which
  we then compute the emissivity and opacities (see also below).

\item[2.] While \texttt{BHAC} advances the conservative RMHD variables
  in time, the LB code does the same for the $N_{\rm pop}$ populations of
  the radiation specific intensity, $I_i$.

\item[3.] After performing the streaming and collision step, the LB code
  computes the zeroth ($E$), first ($F^i$) and second moment ($P^{ij}$)
  of the radiation distribution function according to Eqs.
  (\ref{eq:moments_discrete}) and (\ref{eq:higher_moments}).

\item[4.] From these moments, we compute the radiative source terms
  $\mathcal{S}_0=W(\tilde{\kappa}_aJ-\tilde{\eta})+\tilde{\kappa} H_0$
  and $\mathcal{S}_j=W(\tilde{\kappa}_a J - \tilde{\eta})v_j +
  \tilde{\kappa} H_j$, which we return to \texttt{BHAC}. Here $W$ is the
  Lorentz-factor, $\tilde{\kappa}\coloneqq \tilde{\kappa}_a +
  \tilde{\kappa}_s\coloneqq \tilde{\kappa}_a + (\tilde{\kappa}_0 - 1/3
  \tilde{\kappa}_1)$, and $J$ and $H_j$ are the radiation energy and
  momentum density in the comoving fluid frame, to which we transform via
  \begin{align} \label{eq:J}
    J &= W^2 \big( E - 2F^i \varv_i + P^{ij}\varv_i \varv_j \big) \\
    H_j &= W^3 \big(F^i \varv_i - E\big)\varv_j + Wh_{ji}
    F^i - Wh_{ji} \varv_k P^{ik} \, ,
  \end{align}
  where $h_{ij}=W^2\varv_i \varv_j + \delta_{ij}$ is the projection
  operator orthogonal to the fluid velocity. $H_0$ can then be computed
  from $H_\mu u^\mu = 0$.
\item[5.] After \texttt{BHAC} has updated its variables and before
  cycling to 1), we add the sources $\mathcal{S}_0$ and $\mathcal{S}_j$
  to the energy and momentum equation, respectively, which are solved in
  conservative form within \texttt{BHAC}.

\end{itemize}

As it is the case in most simulations of high-energy astrophysical
phenomena, the fluid moves at relativistic
speeds. Eq. (\ref{eq:rte_gray}), is written in an Eulerian (lab) frame so
that the opacities and emissivities $\eta$, $\kappa_a$, $\kappa_0$,
$\kappa_1$ it employs are to be evaluated in the same Eulerian
frame. However, the microphysics used to derive such quantities is well
defined only in the fluid's rest-frame (\ie the frame co-moving with the
fluid) where the corresponding quantities $\tilde{\eta}$,
$\tilde{\kappa}_a$, $\tilde{\kappa}_0$, $\tilde{\kappa}_1$ are isotropic
and can be written in a compact way. Therefore, care needs to be taken in
transforming the opacities and emissivities between the two frames, as we
discuss in detail in Appendix \ref{sec:appendix2}.

For the setup of our simulation, we follow \citet{Marti97}, who have
extensively analysed relativistic jets in a purely hydrodynamical
context. The jet is simply injected through a circular nozzle at the
lower edge of the computational domain and propagating parallel to the
coordinate $z$-axis in a Cartesian grid. The simulation is then
characterized by four parameters: the Newtonian Mach number
$\mathcal{M}:= v_{\rm jet}/c_s$, with $c_s$ the local sound speed, the
jet Lorentz factor $W:=(1-v_{\rm jet}^2)^{-1/2}$ with $v_{\rm jet}$ the
jet propagation velocity, the ratio of the jet rest-mass density to that
of the ambient medium $\mathcal{R}:=\rho_{\rm jet}/\rho_{\rm amb}$ and
the pressure ratio $\mathcal{K}:=p_{\rm jet}/p_{\rm amb}$. We simulate a
pressure matched jet, \ie $\mathcal{K}=1$, with $\mathcal{R}=0.01$, $W=7$
and $\mathcal{M}=42$ on a grid of size $-7.5 < x,y < 7.5, \, 0 < z < 60$
covered by $160\times160\times640$ grid cells, where the jet is injected
at $z=0$ through a nozzle with radius $r_{\rm jet} = 1$.

Since we are only interested in a proof-of-concept simulation, we limit
ourselves to this simple setup, but refer the reader to \citet{Fromm2018}
for an extension that also includes a non-homogeneous background, as is
to be expected near a gravitational source like a black hole. However, we
include a helical perturbation as in \citet{Aloy1999}, so that the jet
deviates from axisymmetry, leading to a truly three-dimensional structure
that allows us to test all terms of the LB code. In all cases, we
consider the magnetic field to be zero.

For the solution of the RTE within our LB scheme we choose a Lebedev
stencil with $N_{\rm pop}=154$ discrete velocity directions. We
initialise the populations at zero and let the radiation evolve
self-consistently during the simulation. To this purpose, we set
$\tilde{\kappa}_a = \rho^2 T^{-3.5}$ and $\tilde{\eta} = \sigma_{_{\rm
    SB}} / \pi \tilde{\kappa}_a T^{4}$, where the fluid temperature $T$
is computed from an ideal gas equation of state. This absorption opacity
is motivated by the Rosseland mean opacity for thermal bremsstrahlung
\citep{Rybicki_Lightman1986} and the emissivity simply follows from
Kirchhoff's law upon assuming black-body radiation.

We should note that in contrast to the pure-RMHD simulation, which is
scale invariant, the coupled RMHD-radiation simulation fixes the
length scale via the value of the Stefan-Boltzmann constant $\sigma_{_{\rm
    SB}}$ and the density assumed for the ambient medium $\rho_{\rm
  amb}$. Here, we simply use $\rho_{\rm amb}=1$ and $\sigma_{_{\rm
    SB}}=0.1$, which does not lead to a physically realistic setup, but
ensures that a moderate amount of radiation is produced, that neither
dominates the fluid nor is dominated by it.

Finally, we also add scattering using $\tilde{\kappa}_0=10^{-3}\rho$ and
$\tilde{\kappa}_1=0$. This choice is motivated by the microphysical
process of Thomson scattering, which is proportional to the number of
scatters in the medium (hence, the choice for $\tilde{\kappa}_0$) and has
no preferred direction (hence, the choice for $\tilde{\kappa}_1$).

We compare the results of the pure-RMHD and the RTE-coupled simulations
in Fig. \ref{fig:jets}, whose left panel refers to the pure-RMHD jet, the
central panel to the RTE-coupled solution obtained with the LB method,
and the right panel to the corresponding evolution when the RTE-coupled
solution is obtained with the M1 scheme. A quick comparison of the
pure-RMHD jet morphology in the left panel shows that it is in good
agreement with the one presented by \citet{Marti97} and \cite{Aloy1999},
where this type of jets has been studied extensively. On the other
hand, the solutions employing a coupling with the radiation are
considerably different. In particular, the RTE-coupled simulation with the LB
method shows that the jet propagates more slowly, \ie the Lorentz factor
is $\sim 15\%$ smaller than for the pure-RMHD case. This is due to the
fact that the fluid making up the jet loses energy via the emission of
radiation. By stark contrast, the RTE-coupled solution obtained with the
M1 scheme shows that the jet propagates more rapidly. While this behaviour
is similar to the one reported by \citet{RiveraPaleo2019} -- who, in
addition, continuously injected energy in the radiation field -- we
believe it is actually incorrect.

The origin of this substantially different dynamical behaviour is due to
the poor handling by the M1 scheme of radiation interacting with itself.
We find this to cause significantly different distributions of the
energy-density. 
In the case of the M1-evolution, the radiation energy density accumulates mostly
in a narrow region along the $z$-axis, while it is more evenly spread in
the case of LB-evolution. The actual cause of this radiation focussing is
the same as at the origin of the poor performance of the M1 scheme in the
beam-crossing test reported in Fig. \ref{fig:crossing_beams}. Also in
this case, in fact, different beams of radiation originating from the
recollimation shock produced near the injection region of the jet
\citep{Mizuno2015}, intersect along the $z$-axis and lead to an
inconsistent combination of radiation fluxes.

\begin{figure}
  \includegraphics[width=0.99\columnwidth]{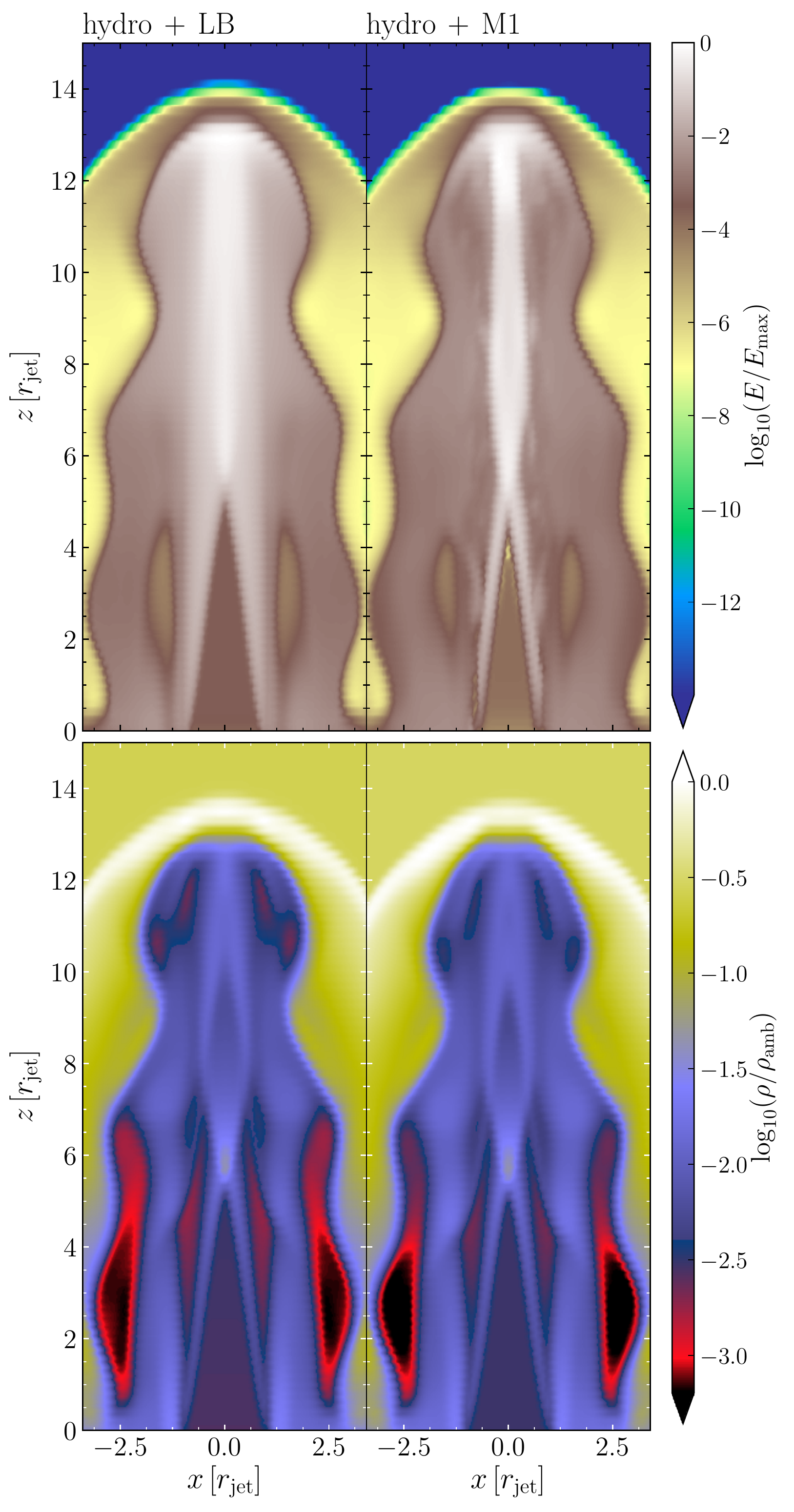}
  \caption{\textit{Top panel:} Cut through the $(x,z)$ plane of the
    radiation energy density at $t=30\,r_{\rm jet}$ for the coupled
    RMHD-radiation simulations using the LB method (left) and the M1
    scheme (right). \textit{Bottom panel:} The same as in the
    top panel, but for the rest-mass density.}
  \label{fig:jets_early}
\end{figure}

This is illustrated in the top panels of
Fig. \ref{fig:jets_early}, which reports a cut through the $(x,z)$ plane
of the radiation energy density $E$ at an early time during the jet
evolution, namely at $t=30\,r_{\rm jet}$, using either the LB method
(left) or the M1 scheme (right).  Note that in both cases there is a
triangular region of small $E$ originating from the recollimation shock,
and ultimately due to the contact discontinuity between propagating jet
and ambient medium \citep[see][for a discussion of the role of the
  contact discontinuity in accelerating the jet]{Aloy:2006rd}. This
region of small $E$ is surrounded on both sides by high-$E$ streams,
which appear as white in the colorcode used. In the case of the
M1-evolution (right panel), these two streams merge at the top of the
triangular low-$E$ region leading to the same pathology discussed in
Sec. \ref{sec:beam_tests} (\cf Fig. \ref{fig:crossing_beams}) and to the
artificial accumulation of radiation along the $z$-axis. The latter then
provides additional momentum, pushing the jet forward and
overcompensating the linear momentum lost in the production of the
radiation. In the LB-evolution, on the other hand, the two beams do not
merge but cross correctly. As a result, the radiation energy density is
not artificially focused and $E$ spreads out over a larger region,
leading to a broader bow shock ahead of the jet, which is also
propagating more slowly.

Note that these problems in the M1-evolution affect the dynamics of the
jet only downstream of the first recollimation shock. This is very
clearly illustrated in the bottom panels of Fig. \ref{fig:jets_early},
which refer instead to the rest-mass densities in the two cases and to
the same time as the top panels, \ie $t=30\, r_{\rm jet}$. As one would
expect, the differences are in this case very small since the radiative
effects have not yet been able to play a role, which they will instead do
for $t \gtrsim 30\, r_{\rm jet}$.

In summary, these results go well-beyond our intention of providing
proof-of-principle evidence for a correct implementation of the LB method
in a fully coupled relativistic-fluid configuration. In particular, they
clearly show the ability of the LB method to handle correctly scenarios
with physical conditions that are very close to those encountered in
relativistic astrophysical phenomena. More importantly, they highlight
that under these very same conditions, the M1 approach commonly employed
-- even by us \citep{Weih2020} -- may suffer from artefacts that may
affect the dynamics of relativistic jets, such as those expected to be
generated in a binary system of merging neutron stars, and whose accurate
description is essential to gain insight in the launching of relativistic
jets in short gamma-ray bursts.

\section{Performance and Scalability}
\label{sec:benchmarks}

Novel computational methods are expected to i) feature good scalability
on parallel computers, ii) perform better than existing ones in accuracy,
efficiency or both.

In what follows we show that the LB method proposed here possesses both
these qualities, i.e. parallel scalability and efficiency.

Indeed, one key ingredient in the success and widespread adoption of LB
is parallel efficiency. Thanks to the synchronous algorithmic flow
stemming from the stream-collide paradigm, LB schemes exhibit high
amenability to parallel coding, making them natural targets for efficient
implementations on modern computer architectures \citep{walberla, muphy,
  hemelb, bernaschi-cc-2010, succi-cf-2019}.

In this section, we provide a brief performance evaluation of the
numerical scheme designed in this paper.
With respect to classical LB models, our scheme presents two main
differences: \textit{i)} we employ off-lattice quadratures, which require
interpolation to implement the streaming phase; \textit{ii)} the number
of velocity components forming the stencil is significantly larger with
respect to standard LB schemes.

\begin{figure}
  \includegraphics[width=0.99\columnwidth]{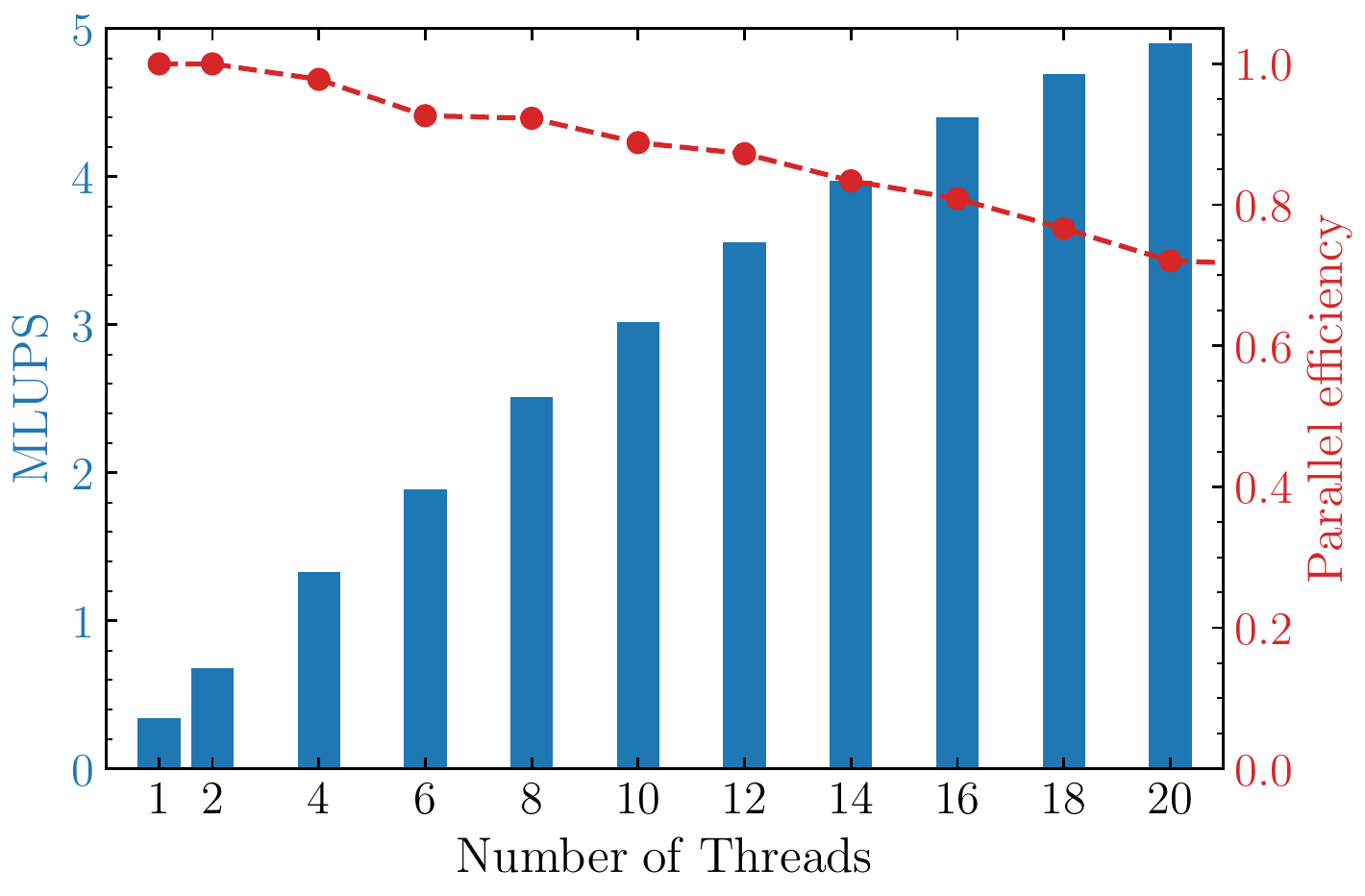}
  \caption{Analysis of performance scalability  on a single Intel
	Skylake board. The results refer to simulations of the emitting-sphere
	benchmark with an explicit Euler stepper, on an $L=128^3$ grid, using a
	stencil with $N_{\rm pop} = 200$ components. The figures reported
	correspond to the best out of 10 trials for each set of parameters. }
  \label{fig:scalability-skylake}
\end{figure}

We have implemented our numerical scheme using directive-based
programming environments, such as \textit{OpenMP} and \textit{OpenACC},
to expose parallelism \citep{calore-cc-2016}. The advantage of this
approach is that the code is portable and can therefore be compiled and
executed on diverse architectures, from commodity CPU-based processors to
GPU accelerators.
We test the code on an Intel Skylake 20-cores processor, a commonly
adopted architecture in the HPC-market. We measure performances in terms
of Million Lattice Updates per Second (MLUPS):
\begin{equation}
  \text{MLUPS} = 10^6 \frac{ L^3 ~ N_{\rm iter} }{t_{\rm exe}} \,,
\end{equation}
where $t_{\rm exe}$ measures the execution time (in seconds) required to
simulate $N_{\rm iter}$ timesteps on a grid of $L^3$ points. In essence,
for a problem of fixed size and iterations $L$ and $N_{\rm iter}$, a
parallel implementation should lead to an execution time that decreases
linearly with the number of cores and hence a linearly growing
MLUPS. This is shown in Fig.~\ref{fig:scalability-skylake}, where we
assess the scalability of the code on a single Intel Skylake board. We
solve the emitting-sphere benchmark with an explicit Euler stepper, on an
$L=128^3$ grid, and with a spherical-design quadrature of order $p=20$
with $N_{\rm pop} = 200$ discrete components. The code scales up to 20
threads with a parallel efficiency above $70 \%$.

These figures are already quite good, but could be further improved.
A limiting performance factor is given by inefficient memory accesses,
which in turn leads to a sub-optimal use of the cache and of the vector
unit of the processor. Omitting details, we simply mention here that a
careful optimisation of the data-layout used to store the grid in memory
should be taken into consideration.
The two most common data layouts used in several stencil applications are
the so-called array of structures (AoS) and structure of arrays (SoA)
schemes; in the AoS layout, all populations associated to one lattice
site are stored in contiguous memory locations. Conversely, in the SoA
scheme all the populations having the same index $i$ are stored
contiguously, while populations belonging to the same lattice site are
stored far from each other at non unit-stride addresses.
Recently, more sophisticated data-layouts have been designed explicitly for LB
applications \citep{shet-pre-2013} yielding strong benefits, especially when
targeting multi-core architectures with wide vector units and large cache
memories \citep{calore-ijfpca-2019}. The application of these optimisation and
a more in-depth analysis will be reported elsewhere.

\begin{figure}
  \includegraphics[width=0.99\columnwidth]{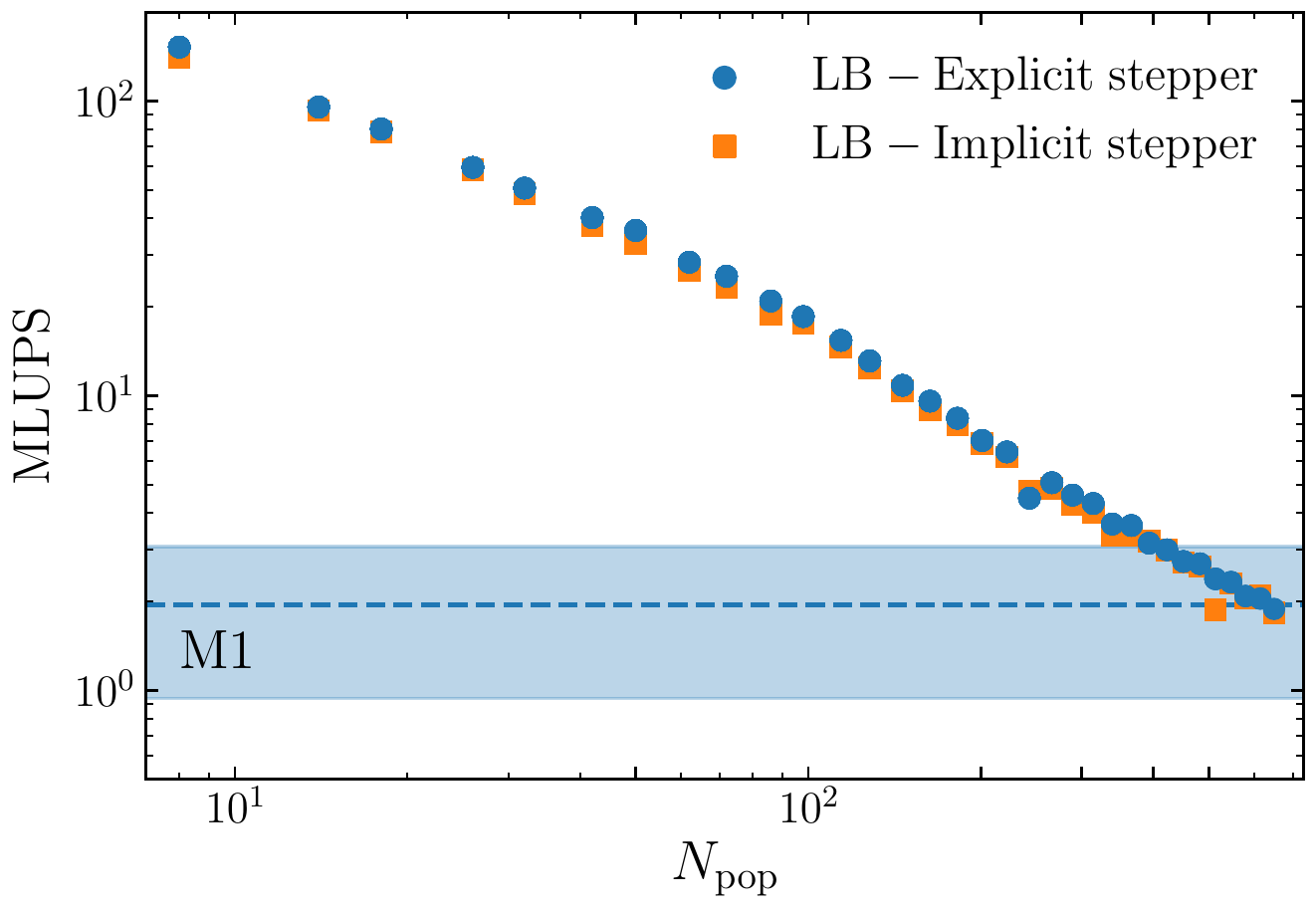}
  \caption{ Comparison of the performance (measured MLUPS) achieved on a
    dual-Skylake processor (40 threads) as a function of $N_{\rm pop}$,
    the number of the discrete components forming the stencil. The
    results refer to simulations of the emitting-sphere benchmark, on an
    $L=128^3$ grid. Dots correspond to simulations based on an explicit
    Euler stepper, while squares correspond to simulations using an
    implicit stepper. The dashed line shows the performance of the M1
    code \texttt{FRAC} with the blue-shaded band corresponding to
    different accuracies for the M1 closure. The figures reported are the
    best out of 10 trials for each set of parameters. }
  \label{fig:performance-cmp}
\end{figure}

In Fig.~\ref{fig:performance-cmp}, we show the performance obtained on a
dual-Skylake processor as a function of $N_{\rm pop}$. We show results
for an explicit and also for the implicit stepper described in
Sec. \ref{sec:numerical}, which extends the stability of the method also
to the case of very stiff terms. To simplify the comparison, we have
executed the \textit{Lambda iteration} loop exactly $5$ times at each
grid-cell. Since the execution time of the current implementation is 
completely bounded by memory accesses, the difference between the implicit 
and the explicit scheme is almost negligible.

The figure also presents a comparison with the performance of the
special-relativistic version of the M1 code \texttt{FRAC}
\citep{Weih2020b}, featuring a similar level of optimisation. For M1 the
major bottleneck is the closure relation, which requires a complex
root-finding algorithm for obtaining the pressure tensor. In principle,
the performance of \texttt{FRAC} can be arbitrarily good or bad
depending on the desired accuracy of the root-finding step. Here we
consider a standard setup (\ie a relative accuracy of $10^{-9}$ for
root-finding and computing the closure on cell-centres as well as
cell-faces). As can be seen from the dashed line in
Fig. \ref{fig:performance-cmp}, the LB method outperforms the M1 code for
$N_{\rm pop} \lesssim 400$, the latter being a rather generous number of
discretized directions to solve most astrophysical problems. We also
consider somewhat higher ($10^{-14}$) and lower ($10^{-4}$) relative
accuracy for the root-finding, which is shown as a blue band in
Fig. \ref{fig:performance-cmp}.

Finally, we comment that the LB method is particularly well suited for
GPU implementations \citep{bernaschi-cc-2010,Calore2017}.
For this reason we also test a GPU-optimized version of our code on an
NVIDIA V100 GPU. We observe the performances to be systematically one
order of magnitude higher than those reported on the CPU, both for
stencils with $10-100$ components, suitable for simulations in which
scattering terms are relevant, as well as for stencils with $>200$
components, which instead allow to extend the applicability of the solver
to free-streaming regimes. The present version of the code allows for
efficient memory accesses on GPUs architectures, in turn exposing 
the higher costs of the implicit solver, which for large values of
$N_{\rm pop}$ is found to be $2-2.5 \times$ slower than the explicit 
Euler method.

\section{Conclusions and Outlook}
\label{sec:outlook}

We have presented an extension of the LB method, which is commonly used
in classical fluid dynamics, to the solution of the RTE in special
relativity, thus making it applicable to three-dimensional simulations of
high-energy astrophysical phenomena.

After implementing the new method in flat spacetime and under the grey
approximation, we analysed its performance in a number of code tests.  In
this way, we have shown that while the LB method performs extremely well
in the diffusion limit, the free-streaming regime represents the most
difficult one to treat accurately. In this regime, a large number of
discrete velocity directions is required since the propagation of
radiation is restricted to these directions only. Nevertheless, even in
this optically thin limit, the LB method proves superior to the commonly
used M1 method, as witnessed by the fact that radiation beams cross
correctly.

This feature is important for astrophysical systems with an accretion
disk and torus, from which radiation is expected to cross and lead to the
phenomenology observed in short gamma-ray bursts. Besides this important
advantage, the new LB method also outperforms the M1 method in the
intermediate regime, between diffusion and free-streaming, mostly because
it does not need to rely on a closure relation to compute higher moments,
as it is the case for the M1 scheme, where the pressure tensor is simply
interpolated via the closure.

The accuracy in the calculation of the moments also depends on the
underlying quadrature, for which we have compared three
possibilities. While in our code tests only minor differences could be
seen among these quadratures, it is mathematically clear that the Lebedev
and spherical-design quadratures are more accurate than the
Gauss-Legendre quadrature commonly used in direct Boltzmann solvers
\citep[see \eg][]{Nagakura2018}.

We have also coupled our new radiation code to the GRMHD-code
\texttt{BHAC} and simulated a relativistic jet, where the radiation
back-reacts dynamically onto the fluid. Not only does this represent the
first such simulation using an LB scheme, but also proves that our new
method is indeed applicable to high-energy astrophysics. Furthermore,
when comparing the corresponding results obtained with the more standard
M1 code \texttt{FRAC}, that employs a moment-based scheme, we have shown
that the LB-method solution does not suffer from the inaccuracies that
plague the M1 method.

Finally, we have also shown that the LB method is faster than the M1
method for a number of discrete directions $N_{\rm pop} \lesssim 400$,
which has been shown to be more than enough for accurately simulating the
diffusion limit and gives reasonable results also in the free-streaming
regime.

Depending on the system, one might need higher accuracy in the latter and
thus $N_{\rm pop} \approx 600-800$ might be necessary. In this case the
LB method is slightly more expensive than the M1 scheme. Considering,
however, the high amenability of LB to GPU implementations, a major
speed-up can certainly be achieved along this line.

While this paper is meant to provide a presentation of the LB method for
the solution of the radiative-transfer equation in computational
astrophysics, a number of improvements are possible, both in terms of
astrophysical applications, and in terms of mathematical and numerical
developments. The former involves a more detailed and realistic
investigation of the role played by radiation in impacting the dynamics
and imaging of astrophysical relativistic jets. The latter necessarily
involves the extension of the method to a general-relativistic framework,
which requires the extension of the streaming-step to curved spacetimes.

Finally, the numerical developments will include the possibility of using
numerical grids with various forms of static and dynamic mesh-refinements.
Also in this case, an adjustment of the streaming-step will be needed to
allow for the streaming from coarse to fine grid cells and vice-versa.
Work along these lines is in progress.

\section*{Acknowledgments}

LRW acknowledges support from HGS-HIRe. AG acknowledges funding by
"Contributo 5 per mille assegnato all'Universit\`a degli Studi di
Ferrara-dichiarazione dei redditi dell'anno 2017". DS has been supported
by the European Union's Horizon 2020 research and innovation programme
under the Marie Sklodowska-Curie grant agreement No. 765048. SS
acknowledges funding from the European Research Council under the
European Union's Horizon 2020 framework programme (No. P/2014-2020)/ERC
Grant Agreement No. 739964 (COPMAT). He also wishes to thank A. Ferrara, P. Mocz,
D. Spergel and J. Stone for illuminating discussions. Support also comes
in part from ``PHAROS'', COST Action CA16214; LOEWE-Program in HIC for
FAIR; the ERC Synergy Grant ``BlackHoleCam: Imaging the Event Horizon of
Black Holes'' (Grant No. 610058). The simulations were performed on the
SuperMUC and SuperMUC-NG clusters at the LRZ in Garching, on the LOEWE
cluster in CSC in Frankfurt, on the HazelHen cluster at the HLRS in
Stuttgart, and on the COKA computing cluster at Universit\`a di Ferrara.

\section*{Data availability}
The data underlying this article will be shared on reasonable request to 
the corresponding author.


\bibliographystyle{mnras}
\interlinepenalty=10000 
\bibliography{aeireferences,local}


\appendix

\section{Link between microscopic and macroscopic parameters}
\label{sec:appendix1}

In this section of the appendix  we perform an asymptotic analysis to link the
microscopic parameters with the macroscopic ones. In particular we consider
the specific limit of zero emission and absorption ($\kappa_a = \eta = 0$), for
which the radiative Lattice Boltzmann equation is shown to recover the
diffusion equation
\begin{equation}
  \partial_t E = D \Delta E \,.
\end{equation}

In Appendix~\ref{app:ce} the link between the diffusive coefficient $D$ and the
scattering parameters ($\kappa_0$, $\kappa_1$) is established analytically
through a Chapman-Enskog expansion \citep{chapman-book-1970}. In
Appendix~\ref{app:num} the analytic expressions are extended to account for
extra numerical corrections by fitting results from numerical simulations.

\subsection{Chapman-Enskog Analysis}
\label{app:ce}

We start from Eq. (\ref{eq:rte_discrete}) assuming zero absorption and
emission, \ie
\begin{equation}\label{eq:rte-discrete}
    I_i(\bm{r} + c\bm{\hat{n}}_i\Delta t, t + \Delta t) - I_i(\bm{r},t) = - c \kappa_0 \Delta t \left( I_i(\bm{r},t) - I_i^{\rm eq}(\bm{r},t) \right) \, .
\end{equation}
with $I_i^{\rm eq}(\bm{r},t)$ given by Eq. (\ref{eq:source-and-equil}). Note
that throughout this section we will write $c$ explicitly and write vector
components with Greek indices rather than using boldface vectors. We also adopt
the Einstein convention of summing over repeated indices.

Taking a Taylor expansion of the left-hand side of Eq. (\ref{eq:rte-discrete}), and
including terms up to the second order, gives:
\begin{align} \label{eq:taylor}
  I_i (\bm{x} + \bm{n}_i \Delta t, t + \Delta t) &- I_i (\bm{x}, t ) 
  = 
                \Delta t   \left( \partial_t + c n_i^{\alpha} \partial_{\alpha} \right) I_i \nonumber \\
  &+ \frac{1}{2} \Delta t^2 \left( \partial_t + c n_i^{\alpha} \partial_{\alpha} \right)^2 I_i
  + \mathcal{O}(\Delta t^3) \, .
\end{align}
We also expand the differential operator with respect to time
\begin{equation}\label{eq:ce-dtexp}
  \partial_t = \epsilon   \partial^{(1)}_t
             + \epsilon^2 \partial^{(2)}_t
             + \mathcal{O}(\epsilon^3) \, ,
\end{equation}
and space:
\begin{equation}\label{eq:ce-dxexp}
  \partial_{\alpha} = \epsilon \partial_{\alpha}^{(1)} +
	\mathcal{O}(\epsilon^2) \, ,
\end{equation}
where $\epsilon \ll 1$.

Next, we expand the specific intensity around its equilibrium:
\begin{equation}\label{eq:ce-fexp}
  I_i = I_i^{(0)} + \epsilon I_i^{(1)} + \epsilon^2 I_i^{(2)} +
	\mathcal{O}(\epsilon^3)\,  ,
\end{equation}
where $I_i^{(0)} \equiv I_i^{\rm eq}$.
We also recall the definition of the first and second moment of the
distribution:
\begin{equation}
  E=\sum_i I_i \, ,
\end{equation}
\begin{equation}
  F^{\alpha}=\sum_i n_i^{\alpha} I_i \, .
\end{equation}
Assuming the most basic level of isotropy for the stencil used in the numerical
method we have
\begin{align}\label{eq:stencil-symmetry}
  &\sum_i w_i = 1                                                          \, , \quad 
   \sum_i w_i n_i^{\alpha} = 0                                             \, , \quad \nonumber \\
  &\sum_i w_i n_i^{\alpha} n_i^{\beta} = \frac{1}{d} \delta_{\alpha \beta} \, , \quad 
   \sum_i w_i n_i^{\alpha} n_i^{\beta} n_i^{\gamma} = 0                    \, ,
\end{align}
where $\alpha, \beta$, and $\gamma$ run over the spatial indexes in $d$ dimensions.

By integrating Eq. (\ref{eq:source-and-equil}), in combination with Eq.
(\ref{eq:stencil-symmetry}), we get the following definitions for the moments
of the equilibrium distribution:
\begin{align}
& \sum_i I_i^{\rm eq} = E \,, \label{eq:feq-o0} \\
& \sum_i I_i^{\rm eq} n^{\beta}_i 
  = 
  \frac{\lambda}{d} F^{\beta} \,, \label{eq:feq-o1} \\
& \sum_i I_i^{\rm eq} n^{\beta}_i n^{\gamma}_i
  =
  \frac{1}{d} \delta_{\beta \gamma} E \,. \label{eq:feq-o2} 
\end{align}
Since we are neglecting absorption and emission and consider the diffusion
limit, where the radiation is in thermodynamic equilibrium with the underlying
fluid, we find due to conservation
\begin{equation}
  \int (I - I^{\rm eq}) d \Omega = 0  \Rightarrow \int I d \Omega = \int I^{\rm eq} d \Omega \,,
\end{equation}
which in the discretized form leads to the condition
\begin{equation}\label{eq:ce-consO0}
  \sum_i I_i = \sum_i I_i^{\rm eq} = E \, , \quad \sum_i I_i^{(k)} = 0 \quad \forall k \geq 1 \,.
\end{equation}
We do not write down any condition for the first moment, since its preservation
depends on the choice of $\lambda$.

Replacing the left-hand side of Eq. (\ref{eq:rte-discrete}) with its second order Taylor
expansion, \ie Eq. (\ref{eq:taylor}), we get
\begin{align}\label{eq:rte-discrete-taylor}
  &                \Delta t   \left( \partial_t + c n^{\alpha}_i \partial_{\alpha} \right)   I_i (\bm{x}, t )
  + \frac{1}{2} \Delta t^2 \left( \partial_t + c n^{\alpha}_i \partial_{\alpha} \right)^2 I_i (\bm{x}, t )
  \nonumber \\ 
  &= 
  - c \Delta t \kappa_0 \left( I_i (\bm{x}, t ) - I^{\rm eq}_i (\bm{x}, t )
	\right) \, .
\end{align}
We now plug in the above Eqs. (\ref{eq:ce-dtexp}), (\ref{eq:ce-dxexp}), and
(\ref{eq:ce-fexp}) and perform a multi-scale expansion in which we keep
track separately of terms up to order $\epsilon$ and $\epsilon^2$. The
resulting equations are
\begin{align}
  \mathcal{O}(\epsilon  ) : &  \left( \partial^{(1)}_t + c n^{\alpha}_i \partial_{\alpha}^{(1)}  \right) I_i^{(0)} 
                               \approx 
                               - c \kappa_0 I_i^{(1)} \, , \label{eq:ce-eps01}\\
  \mathcal{O}(\epsilon^2) : &  \left( 1 - \frac{\Delta t}{2} c \kappa_0 \right)
                               \left( 
                                      \partial^{(1)}_t + c n^{\alpha}_i \partial_{\alpha}^{(1)}  
                               \right) I_i^{(1)} 
                               +  
                               \partial^{(2)}_t I_i^{(0)} 
                               \approx
                               - c \kappa_0 I_i^{(2)} \, . \label{eq:ce-eps02}
\end{align}

We then take into consideration Eq. (\ref{eq:ce-eps01}) and integrate (i.e. sum
over all $N_{\mathrm{pop}}$ populations), getting
\begin{equation}
  \sum_i \left( \partial^{(1)}_t I_i^{(0)}  + c n^{\alpha}_i \partial_{\alpha}^{(1)} I_i^{(0)} \right)
  \approx 
  - c \kappa_0 \sum_i I_i^{(1)} \, ,
\end{equation}
which using Eq. (\ref{eq:ce-consO0}) and the definition of the first order
moment leads to
\begin{equation}
  \partial^{(1)}_t E + c \frac{\lambda}{d} \partial_{\alpha}^{(1)} F_{\alpha} =
	0 \, .
\end{equation}

Next, starting again from Eq. (\ref{eq:ce-eps01}), we multiply by $n^{\beta}_i$
and integrate, which yields
\begin{equation}\label{eq:non-conserving-o1}
  \frac{\lambda}{d}  \partial^{(1)}_t F^{\beta} + \frac{1}{d} c
	\partial_{\beta} E = - c \kappa_0 \sum_i n_i^{\beta} I_i^{(1)} \, .
\end{equation}
Note that the RHS of the above equation vanishes when the conservation of the
first moment is ensured; for the moment we leave it in a general form and
evaluate this term later on.

We now integrate Eq. (\ref{eq:ce-eps02}) and obtain
\begin{align}
  & \sum_i
        \left( 1 - \frac{\Delta t}{2} c \kappa_0 \right)
        \left( \partial^{(1)}_t + c n^{\alpha}_i \partial_{\alpha}^{(1)} \right)  I_i^{(1)}
     +  
  \sum_i
     \partial^{(2)}_t I_i^{(0)} 
  \approx
  - \kappa_0 \sum_i I_i^{(2)} \, , \\
  &  \left( 1 - \frac{\Delta t}{2} c \kappa_0 \right)
     c \partial_{\alpha}^{(1)} \sum_i n^{\alpha}_i I_i^{(1)}
     +  
     \partial^{(2)}_t E
  = 0 \, , \\
  & \partial^{(2)}_t E
  =
  - \left( 1 - \frac{\Delta t}{2} c \kappa_0 \right)
     c \partial_{\alpha}^{(1)} \sum_i n^{\alpha}_i I_i^{(1)} \, .
\end{align}
The RHS can be derived from Eq. (\ref{eq:non-conserving-o1}), leading to
\begin{align}
  & \partial^{(2)}_t E
  =
  \frac{1}{c \kappa_0} \left( 1 - \frac{\Delta t}{2} c \kappa_0 \right)
     c \partial_{\alpha}^{(1)} \left( \frac{\lambda}{d}  \partial^{(1)}_t
	F^{\alpha} + \frac{1}{d} c \partial_{\alpha} E \right) \, ,
\end{align}
which can be re-arranged as
\begin{equation}\label{eq:ce-diffusive-eq}
  \partial^{(2)}_t E 
  =
  D
  \left(
  \Delta^{(1)} E
  +
  \lambda \frac{1}{c} 
  \partial_{\alpha}^{(1)}
  \partial^{(1)}_t F_{\alpha}
  \right)
\end{equation}
with
\begin{equation}\label{eq:numerical-diff-coeff}
  D
  =
  \frac{c^2}{d c \kappa_0} \left( 1 - \frac{\Delta t}{2} c \kappa_0 \right)  \,.
\end{equation}
We note that Eq. (\ref{eq:ce-diffusive-eq}) is the diffusion equation with $D$
its diffusion coefficient plus an extra error term. This error term is the same
one that arises for the equations of the M1 system, when considering the
optically thick limit and neglecting absorption and emission. It is remarkable
that despite the M1 system being a set of macroscopic equations, where the
connection between $D$ and the system's scattering coefficient is directly
evident, in contrast to the microscopic equation of our LB method, we arrive at
the exact same macroscopic equation for the diffusion limit.

\subsection{Numerical Fit}
\label{app:num}

In the previous section we have established a link between the microscopic
parameters and the diffusion coefficient by performing the Chapman-Enskog
expansion. In principle one would need to extend the expansion to include
corrections coming from higher order terms. Besides, one should also account
for extra dissipative effects coming from the fact that the LB method described
in the present work is not based on a space-filling Cartesian lattice and
consequently requires interpolation.
Since the overall analysis would become rather tedious from an analytical point
of view, in this section we numerically evaluate the corrections to Eq.
(\ref{eq:numerical-diff-coeff}) that so far have been neglected.

\begin{figure}
  \includegraphics[width=0.99\columnwidth]{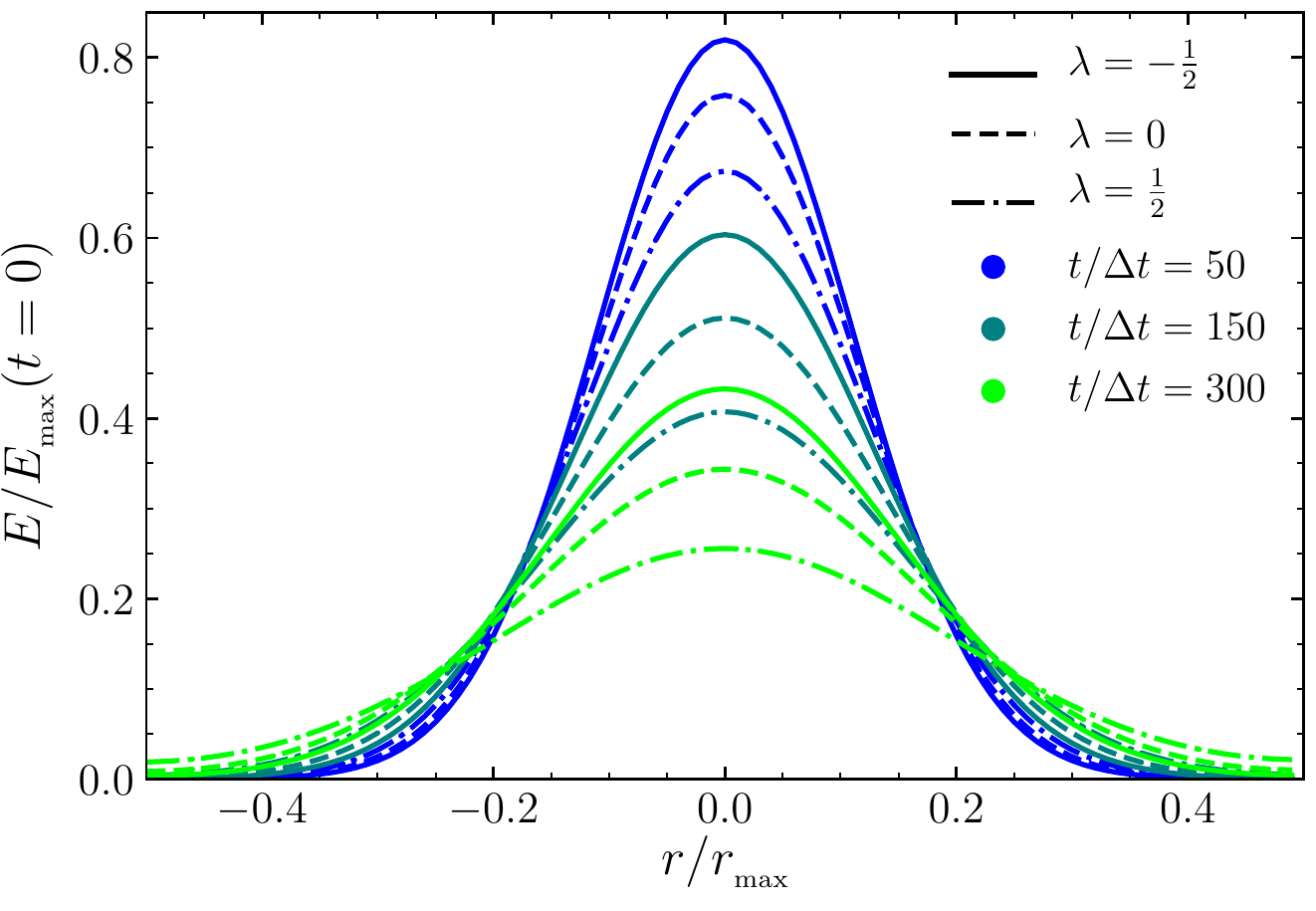}
  \caption{Effect of the parameter $\lambda$ on the diffusion speed of a
    Gaussian pulse. The simulations are performed in a bidimensional
    grid of size $L=100$, with $\Delta t = 1$ and $k_0 \Delta x = 1$. }
  \label{fig:lambda-effect}
\end{figure}

We start by taking into consideration the effect of varying the parameter
$\lambda$. We consider the same numerical setup discussed in
Sec.~\ref{sec:scattering}, working in 2D, on a $100 \times 100$ grid,
with $\Delta t = \Delta x$ and $\kappa_0 \Delta x = 1$. In Fig.
\ref{fig:lambda-effect} we show the results of numerical simulations for
a few selected values of $\lambda$; the results clearly show how
$\lambda$ impacts the diffusion speed.
Since Eq. (\ref{eq:numerical-diff-coeff}) does not depend on $\lambda$, but
only on $\kappa_0$, we extend it by assuming a dependency on the parameter
\begin{equation}
  \chi = \kappa_0 \left( 1 + \alpha_1  \frac{1}{d} \lambda \right) \, , 
\end{equation}
\ie{} a linear combination of $\kappa_0$ and $\kappa_1$, with $\alpha_1$ a
coefficient left to be determined.

In Sec.~\ref{sec:free-streaming} we have discussed the artificial diffusivity
introduced by the interpolation scheme used to implement the propagation step.
In order to try to capture these effects we propose the following extension for
Eq. (\ref{eq:numerical-diff-coeff}):
\begin{equation}\label{eq:numerical-diff-coeff2}
  D
  =
\frac{c}{d \chi } \left[ 1 - \left(\frac{1}{2} + \alpha_2 \right)
\Delta t c \chi \right] + \frac{\Delta x^2}{\Delta t} \alpha_3 \, .
\end{equation}
In the above, $\alpha_2$ introduces a correction to the leading term of
the Taylor expansion of the propagation step, which is also present in
other off-grid LB schemes (see e.g. \cite{coelho-cf-2018}). This
correction becomes vanishingly small as one takes smaller timesteps. We
attribute this correction to the specific interpolation scheme employed.
Besides, we also introduce an extra coefficient $\alpha_3$ which introduces a
background diffusivity which depends on the specific stencil taken into
consideration.

\begin{figure}
  \includegraphics[width=0.99\columnwidth]{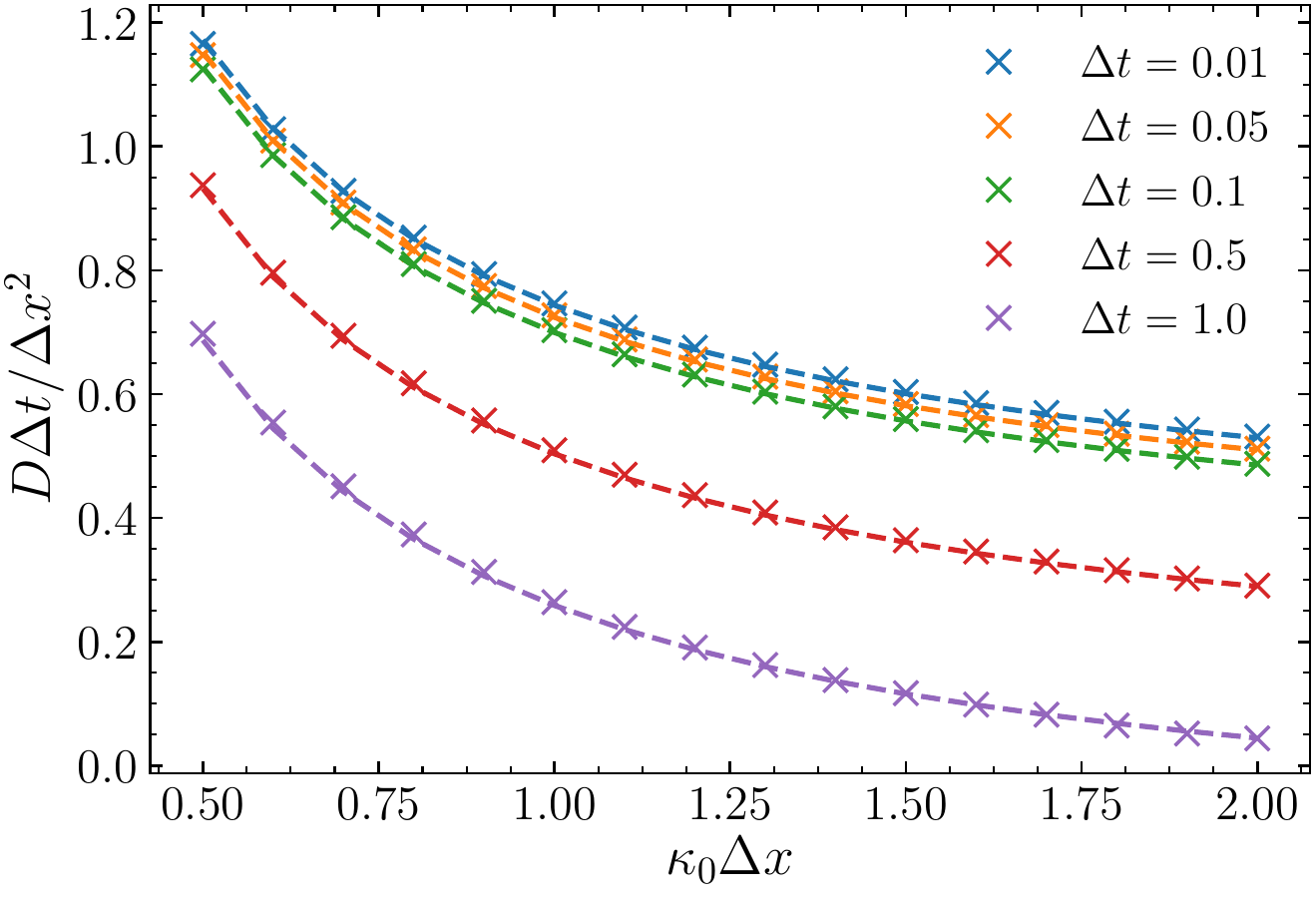}
  \caption{Example of the numerical analysis performed to fit the expression 
           for the diffusion coefficient as in Eq. (\ref{eq:numerical-diff-coeff2}).
           The results are shown for the specific case $\lambda = 0.25$.
           Crosses represent estimates of the numerical diffusion coefficient
           which ensures that the L2-norm of the relative error is within
           $1\%$ with respect to the analytic solution. Dashed lines show the 
           predicted diffusion coefficient using Eq. (\ref{eq:numerical-diff-coeff2})
           with $\alpha_1 = 3/4$, $\alpha_2 = 0.48$ and $\alpha_3 = 0.65$.
           }
  \label{fig:fit-example}
\end{figure}

Our task consists now in determining $\alpha_1, \alpha_2, \alpha_3$.
To this aim we perform several numerical simulations, in which we once
again reproduce the benchmark described in Sec.~\ref{sec:scattering},
varying $\kappa_0 \Delta x$,  $\Delta t$ and $\lambda$.
We follow the evolution of each simulation up to $t = 200 \Delta t$.
Next, we estimate the numerical diffusion coefficient of each simulation.
To do so we compare the numerical results with the analytic solution 
(Eq. (\ref{eq:gauss-hill-analytic}) ), and fit a numerical value of the diffusion
coefficient which leaves the L2-norm of the relative error below $1\%$ over
several finite time-steps ($t/ \Delta t = 5, 10, 20, \dots 200$).
The crosses in Fig~\ref{fig:fit-example} provide an example of the results
obtained following this procedure, for the specific case $\lambda = 0.25$.

Finally, we attempt to fit the dataset with Eq.
(\ref{eq:numerical-diff-coeff2}).
We find that to good accuracy $\alpha_1 \approx - \frac{4}{3}$ in both 2 and 3
dimensions.
The parameters $\alpha_2$ and $\alpha_3$, however, depend both on the
dimensionality and on the specific stencil taken into consideration, although
both tend to stabilise when considering stencils formed by a sufficiently large
number of components.
To give an example, conducting the analysis in 2D with $N_{\mathrm{pop}}
= 120$ we get $\alpha_2 \approx 0.48$, $\alpha_3 \approx 0.65$. In 3D,
instead, using a spherical-design quadrature with order $p=20$ and
$N_{\mathrm{pop}} = 222$, we obtain $\alpha_2 \approx 0.58$, $\alpha_3
\approx 0.77$.

\section{Emissivity and opacities in the lab frame}
\label{sec:appendix2}

For radiative transfer simulations on a moving fluid background it is simpler
to express the microscopic quantities describing emission, absorption and
scattering processes in the comoving fluid-frame, in which the fluid is at
rest. Since our LB scheme is designed in the lab frame, it is then necessary to
transform these fluid-frame quantities to their lab-frame counterparts. For
doing so we follow the derivation in \citet{Mihalas2001}, which for
completeness we summarise here for the grey approximation.

For a fluid moving with three-velocity $\varv_i$ and Lorentz factor $W$, 
the frequency $\nu$ in the lab frame of a radiation
particle propagating in direction $\hat{n}_i$ is transformed to the comoving
fluid frame via
\begin{equation}
	\tilde{\nu} = W\nu (1-\varv_i \hat{n}^i) \, .
\end{equation}
It was then shown that the frequency-dependent quantities, which are
isotropic in the fluid frame, transform like
\begin{align}
	\eta_\nu &= \frac{\nu^2}{\tilde{\nu}^2}\, \tilde{\eta}_{\tilde{\nu}} =
	\frac{\tilde{\eta}_{\tilde{\nu}}}{W^2(1-\varv_i\hat{n}^i)^2} \\
	\kappa_{a, \nu} &= \frac{\tilde{\nu}}{\nu}\, \tilde{\kappa}_{a,\tilde{\nu}}
	= W(1-\varv_i\hat{n}^i) \tilde{\kappa}_{a,\tilde{\nu}} \, .
\end{align}
The frequency-averaged absorption opacity defined by Eq. (\ref{eq:kappa})
then simply follows as
\begin{equation}
	\kappa_a = \frac{\int_0^\infty W(1-\varv_i\hat{n}^i)
	\tilde{\kappa}_{a, \tilde{\nu}} I_\nu \, d\nu}
	{\int_0^\infty I_\nu \, d\nu} = W(1-\varv_i\hat{n}^i)
	\tilde{\kappa}_a \, ,
\end{equation}
where $\tilde{\kappa}_a$ is the frequency-integrated fluid-frame absorption
opacity.
The transformation of the frequency-integrated emissivity is
\begin{align} \label{eq:eta_l2f}
	\eta = \int_0^\infty \nu^3 \eta_\nu \, d\nu 
	= \frac{\int_0^\infty \tilde{\nu}^3 \tilde{\eta}_{\tilde{\nu}}\, d\nu}{W^2(1-\varv_i\hat{n}^i)^2} 
       &= \frac{\int_0^\infty \tilde{\nu}^3 \tilde{\eta}_{\tilde{\nu}}\,
	d\tilde{\nu}}{W^3(1-\varv_i\hat{n}^i)^3} \\ \nonumber
       &=\frac{\tilde{\eta}}{W^3(1-\varv_i\hat{n}^i)^3} \, ,
\end{align}
where $\tilde{\eta}$ is the frequency-integrated fluid-frame emissivity. 

The scattering opacities are more complicated to transform and we here only
consider iso-energetic isotropic scattering like the Thomson scattering
process, which we use in Sec. \ref{sec:coupled}. We then recognize that for
$\kappa_0$ we have two terms on the RHS of Eq. (\ref{eq:rte_gray}). The one
proportional to $I$ acts like the absorption term and transforms
correspondingly, \ie 
\begin{equation}
	\kappa_0 I = W(1-\varv_i \hat{n}^i)\tilde{\kappa}_0 I \, .
\end{equation}
The second term proportional to $E$ acts like an emission term, for which we
use the same transformation that we used in Eq. (\ref{eq:eta_l2f}). We then
find
\begin{equation}
	\kappa_0 E = \frac{\tilde{\kappa}_0 J}{W^3(1-\varv_i\hat{n}^i)^3} \, , 
\end{equation}
where $J$ is the radiation energy density in the fluid frame, which can be
computed from the lab-frame moments according to Eq. (\ref{eq:J}).

Taking all of the above transformations together means that for the simulation of the
relativistic jet in Sec. \ref{sec:coupled}, we solve the mixed-frame equation
\begin{equation}
  \frac{1}{c}\frac{\partial I}{\partial t} + \bm{\hat{n}} \cdot \nabla I
  = -W(1-\varv_i \hat{n}^i)(\tilde{\kappa}_a + \tilde{\kappa}_0) I +
	\frac{\tilde{\eta}+\tilde{\kappa}_0 J}{W^3(1-\varv_i\hat{n}^i)^3}
\end{equation}
instead Eq. (\ref{eq:rte_gray}).

\end{document}